\documentclass[11pt,a4paper]{article}
\usepackage{imports}
\usepackage{authblk}

\newcommand{\poly}{\operatorname{poly}}

\newcommand{\E}{\operatorname{E}}

\def\Bigbar#1{\mathrel{\left|\vphantom{#1}\right.\n@space}}

\DeclareMathOperator{\dist}{dist}

\newcommand{\Var}{\mathrm{Var}}
\newcommand{\NN}{\bold{NN}}
\newcommand{\CH}{\dist_{\mathrm{CH}}}
\newcommand{\Ot}{\tilde{O}}

\newcommand{\argmin}{\mathrm{arg min}}

\let\oldvec\vec \renewcommand{\vec}[1]{\oldvec{\mkern0mu#1}}

\makeatletter
\DeclareRobustCommand{\cev}[1]{%
  {\mathpalette\do@cev{#1}}%
}
\newcommand{\do@cev}[2]{%
  \vbox{\offinterlineskip
    \sbox\z@{$\m@th#1 x$}%
    \ialign{##\cr
      \hidewidth\reflectbox{$\m@th#1\vec{}\mkern4mu$}\hidewidth\cr
      \noalign{\kern-\ht\z@}
      $\m@th#1#2$\cr
    }%
  }%
}
\makeatother

\NewDocumentCommand{\fancydef}{O{} O{} m m}{%
    \expandafter\def\csname #3\endcsname{\hyperref[def:#3]{\textcolor{black}{#4}}\xspace}%
    \IfNoValueTF{#1}{%
        \expandafter\def\csname #3s\endcsname{\hyperref[def:#3]{\textcolor{black}{#4s}}\xspace}%
    }{%
        \expandafter\def\csname #1\endcsname{\hyperref[def:#3]{\textcolor{black}{#2}}\xspace}%
    }%
}

\def\compactify{\itemsep=0pt \topsep=0pt \partopsep=0pt \parsep=0pt}

\begin{document}
\title{Fully Dynamic Algorithms for Chamfer Distance}
\date{}
\author[1]{Gramoz Goranci}
\affil[1, 3]{Faculty of Computer Science, University of Vienna, Austria}
\author[2]{Shaofeng H.-C. Jiang\thanks{This research was funded in part by a national key R\&D program of China No. 2021YFA1000900,
and the Austrian Science Fund (FWF) 10.55776/ESP6088024.}}
\affil[2, 5]{Peking University, China}
\author[3]{Peter Kiss\thanks{This research was funded in whole or in part by the Austrian Science Fund (FWF) 10.55776/ESP6088024.}}
\author[4]{Eva Szilagyi}
\affil[4]{
UniVie Doctoral School Computer Science DoCS, Faculty of Computer Science, University of Vienna, Austria}
\author[5]{Qiaoyuan Yang}

\maketitle

\begin{abstract}
 We study the problem of computing Chamfer distance in the fully dynamic setting, where two set of points $A, B \subset \mathbb{R}^{d}$, each of size up to $n$, dynamically evolve through point insertions or deletions and the goal is to efficiently maintain an approximation to $\CH(A,B) = \sum_{a \in A} \min_{b \in B} \textrm{dist}(a,b)$, where $\textrm{dist}$ is a distance measure. Chamfer distance is a widely used dissimilarity metric for point clouds, with many practical applications that require repeated evaluation on dynamically changing datasets, e.g., when used as a loss function in machine learning. In this paper, we present the first dynamic algorithm for maintaining an approximation of the Chamfer distance under the $\ell_p$ norm for $p \in \{1,2
\}$.
 Our algorithm reduces to approximate nearest neighbor (ANN) search with little overhead. Plugging in standard ANN bounds, we obtain $(1+\epsilon)$-approximation in $\tilde{O}(\epsilon^{-d})$ update time and $O(1/\epsilon)$-approximation in $\tilde{O}(d n^{\epsilon^2} \epsilon^{-4})$ update time.
 We evaluate our method on real-world datasets and demonstrate that it performs competitively against natural baselines.

 \end{abstract}

\section{Introduction}

\label{sec:intro}
We consider the problem of computing the Chamfer distance, a popular dissimilarity metric between point clouds. Given two sets of points $A,B \subset \mathbb{R}^d$, each of size up to $n$, the Chamfer distance of $A$ from $B$ with respect to a distance measure $\dist : \mathbb{R}^d \times \mathbb{R}^d \rightarrow \mathbb{R}_{\geq 0}$ is defined as \[ \CH(A,B) = \sum_{a \in A} \min_{b \in B} \dist(a,b).\]
The Chamfer distance is typically defined with respect to distance measures such as Manhattan and Euclidean metrics. It has found a wide range of applications in various domains, including machine learning \cite{KusnerSKW15, WanCLYZY019}, computer vision \cite{AthitsosS03, LiangJG05, FanSG17, JiangSQJ18}, and geometric computing \cite{HarenstamNielsenSSAC24}. Due to its strong empirical performance, the Chamfer distance is often used as a computationally efficient alternative to the more demanding Earth-Mover's distance (EMD) \cite{KusnerSKW15, AtasuM19}.

In many practical applications, the Chamfer distance is repeatedly calculated on evolving datasets. A notable example is cloud completion and up-scaling, where the models aim to reconstruct missing regions or enhance the resolution of 3D point clouds. In such tasks, the Chamfer distance is commonly used as a loss function during training and it must be evaluated continously as the model's predictions evolve \cite{lin2023hyperbolic, LinLZHYFLZZ24, WuPZWLL21}. Two other mainstream use cases include (1) object reconstruction from video sequences, where the objective is to represent objects as point clouds based on observations from a moving camera~\cite{RunzLTMKS0ASLN20, TyszkiewiczMPF22, HampaliHTMKL23}; and (2) medical imaging, where it is used  to track anatomical structures (such as heart motion on ultrasound images) over time~\cite{van1994automatic, herman1991shape, mathews2014segmentation}.

Motivated by these applications, we pose the following fundamental question: \emph{Can the Chamfer distance be maintained under dynamically evolving point sets?} More concretely, consider two input sets $A$ an $B$ that undergo point insertions or deletions, referred to as \emph{updates}. The goal is to design a dynamic algorithm that \emph{efficiently} supports these updates while maintaining an estimate that approximates the Chamfer distance up to a small relative error. A naive solution to handle such updates is to recompute the Chamfer distance from scratch after each update. However,  this is computationally prohibitive:  the best-known static algorithms require either (1) $O(n^2 \cdot d)$ time for exact computation or (2) $O(n d \log n \epsilon^{-2})$~\cite{bakshi2023near}, for a $(1+\epsilon)$-approximation when the underlying metric is an $\ell_p$ norm for $p \in \{1,2 \}$. In summary, even for low-dimensional datasets, these off-the-shelf static algorithms cannot go beyond the linear time barrier for handling updates.

In this paper, we obtain the first dynamic algorithm for maintaining an estimate to the Chamfer distance under the $\ell_p$ norm for $p \in \{1,2\}$, which significantly outperforms the linear-time update barrier. Our problem reduces to nearest neighbour (NN) oracles: given parameters $\alpha > 0, \tau \geq 1$, there is a data structure that maintains a dynamic point-set $B \subset \mathbb{R}^{d}$ and supports in $\tau$ time the following opertions (i) insert/delete a point in $B$ and (ii) given a point $x$, return an $(1+\Theta(\alpha))$-approximate nearest neighbour of $x$ in $B$. We call such a data structure an $(1 + \Theta(\alpha))$-approximate NN oracle with time parameter $\tau$. Plugging in known bounds for NN oracles, our algorithm achieves constant (or even $(1 + \epsilon)$) approximation and supports very fast updates across different parameter regimes. The guarantees of our algorithmic reduction are summarized in the theorem below.

\begin{theorem}
    \label{thm:main}
    Let \(A,B\) be two set of points from \(\mathbb{R}^d\), with $|A|, |B| \leq n$, and let $\epsilon \in (0,1)$ $\alpha > 0$, and $\tau \geq 1$ be parameters. Assume that there is an $(1+\Theta(\alpha))$-approximate NN oracle with time parameter $\tau$. Then there is a dynamic algorithm that supports insertions and deletions of points to $A$ and $B$ in $\tilde{O}(\tau)$ worst-case time per update, and when queried, with high probability, it returns a $(1+ \alpha + \epsilon)$-approximation to $\CH(A,B)$ in $\tilde{O}((d + \tau) \epsilon^{-2} \max\{1,\alpha^2\})$ time, when the underlying metric is the Manhattan $(\ell_1)$ metric.  
\end{theorem}

 The state-of-the-art trade-offs for $(1+\Theta(\alpha))$-approximate NN oracles differ between low and high dimensions; (a) for low dimensions, the trade-off is $\alpha = \epsilon$ with query/update time $\tau = \tilde{O}(\epsilon^{-d})$ ~\cite{AryaMNSW98}, while (b) for high dimensions, we have $\alpha = O(1/\epsilon)$ with $\tau = \tilde{O}(d n^{\epsilon^{2}})$~\cite{AndoniR15}. Substituting these bounds in Theorem~\ref{thm:main}, we obtain dynamic algorithms for the Chamfer distance achieving $(1+\epsilon)$-approximation in $\tilde{O}(\epsilon^{-d})$ update time and an $O(1/\epsilon)$-approximation in $\tilde{O}(d n^{\epsilon^2} \epsilon^{-4})$ update time. \footnote{The stated bounds on the update time are in fact stronger in that they achieve a better dependency on the dimension $d$. This is because Theorem~\ref{thm:main} presents only a simplified version of our main result; the precise trade-offs are given in Theorem~\ref{thm:main_const_prob}.} Moreover, using known embeddings from $\ell_2$ into $\ell_1$, Theorem~\ref{thm:main} readily extends to the setting when the underlying metric is Euclidean ($\ell_2$)~(see Appendix~\ref{app:embeding} for details).
 


Our dynamic algorithms maintain an \emph{estimate} to the Chamfer distance $\CH(A,B)$ with provable approximation ratios. It is natural to ask whether it is possible to maintain the underlying \emph{assignment} $g : A \rightarrow B$ that attains these approximation ratios, i.e., $\sum_{a\in A} \dist(a,g(a)) \leq (1+\epsilon)\cdot \CH(A,B)$. Since reporting the assignment itself takes $\Theta(n)$ time, reporting the \emph{changes} in the assignment due to an update, also known as \emph{recourse}, would be desirable. Unfortunately, it turns out that any dynamic algorithm that maintains an $\alpha$-approximate assignment between $A$ and $B$ must have at least $\Omega({n})$ recourse, and thus also $\Omega({n})$ update time (see \Cref{lem:highRecourse}). In fact, for any constant $\delta > 0$, ~\cite{bakshi2023near} show that even in the \emph{static} setting, reporting a $(1+\epsilon)$-approximate assignment requires $\Omega(n^{2-\delta})$ time under the hitting set conjecture~\cite{williams2018some}.

\paragraph{Experiments.}
We implement our algorithm and validate its performance over four real datasets covering both high and low dimensions, as well as their noisy versions with injected outlier points.
In all datasets, our algorithm achieves less than $10\%$ error using only hundreds of samples,
even against injected outliers,
in time up to magnitudes better than a naive dynamic algorithm.
We also discover that a simple uniform sampling baseline is competitive for these real datasets,
but its performance degrade significantly when outliers are present,
where our algorithm has a clear advantage.

\subsection{Technical Contribution}
We employ an importance sampling framework to estimate $\CH(A, B)$, and
our main contribution is a dynamic data structure that maintains an importance sampler.
This importance sampling framework was introduced in~\cite{bakshi2023near} in a static setting,
where the key idea is to compute a coarse $O(\log n)$-approximate assignment $\widehat{g}$ from $A$ to $B$,
and then sample with probability proportional to the distance $\dist(a, \widehat{g}(a))$.
A standard importance sampling argument shows that an average of $\Ot(1)$ samples suffices for $(1 + \epsilon)$-approximation to $\CH(A, B)$.

However, it is difficult to \emph{explicitly} maintain the approximate assignment $\widehat g$ in the dynamic setting, since the recourse/change of $\widehat g$ can already be very significant per update (let alone the running time).
To resolve this issue, our dynamic sampler obtains these sampling guarantees with an \emph{implicit} representation of distance estimates $\hat{\dist}_a : a \in A$ such that $\hat{\dist}_a$ is an $O(\log^ 2n)$ approximation to $\min_{b \in B}\dist(a,b)$ for $a \in A$ in $\tilde{O}(d)$ update time.
This sampler only generates a sample $\widehat a \in A$, and the final estimator for the importance sampling is computed through $O(\log^2 n \cdot \epsilon^{-2}\max\{1,\alpha^2\}) = \tilde{O}(\epsilon^{-2} \max\{1,\alpha^2\})$ queries to an $(1+\Theta(\alpha))$-approximate nearest neighbor oracle.
This eventually leads to a $(1 + \alpha+\epsilon)$-approximation to $\CH(A,B)$.

Our dynamic sampler (similarly to the static algorithm of \cite{bakshi2023near}) relies on a family of partitions of the input plane $\mathbb{R}^d$ into a series of nested cells that exponentially decrease in size on lower levels (also known as randomly-shifted quadtree in the literature).
For each $a \in A$, we say that \(a\) is matched in some cell of our decomposition if it is the smallest cell that contains $a$ and any point of $B$. The size of this unique sub-cell serves as an approximation to $\min_{b \in B} \dist(a,b)$. Our goal is to implicitly maintain some information about the unique matching cell for all $a \in A$. 

Importantly, we cannot afford to explicitly maintain the matching cell for each point $a \in A$ as it could change for $\Omega(n)$ points of $A$ due to a single update in $B$. Instead, for each cell we maintain how many points of $A$ happen to be matched to $B$ inside that sub-cell. This allows us to implement a sampler that, instead of explicitly sampling a point of $A$, it samples a cell in our nested decomposition based on its size and the number of points of $A$ matched in it. Once a cell is sampled, our goal is to sample a uniformly random point of $A$ matched in that cell. To achieve this, every cell maintains a dynamic sampler which allows it to sample one of its sub-cells in our family of partitioning with probability proportional to the number of points of $A$ in that sub-cell. Repeating this sub-cell sampling process through the $\tilde{O}(1)$ levels of the algorithm finds a cell that contains a single point of $A$, which we then return.



\subsection{Related Work} 
 In the static setting, \cite{bakshi2023near, feng2025even} present a near-optimal algorithm for estimating the Chamfer distance, running in $\tilde{O}(n  d \cdot \epsilon^{-2})$ time. For comparison, our dynamic algorithm handles point updates in near-optimal time proportional to $\tilde{O}(d)$, up to the cost of invoking the nearest neighbour oracle.

Perhaps the closest problem to the dynamic Chamfer distance is the dynamic maintenance of the Earth Mover distance (EMD), for which Chamfer distance is often used as a proxy in practical applications~\cite{KusnerSKW15, AtasuM19}. For $d=2$, dynamic EMD is known to admit an algorithm that achieves $O(1/\epsilon)$ approximation in $O(n^{1/\epsilon})$ update time~\cite{goranci2025fully}. In contrast, our dynamic algorithm for the Chamfer distance extends to \emph{any} dimensions, and can even achieve an improved approximation ratio of $(1+\epsilon)$ in low dimensions. Similar to our negative result on the recourse, ~\cite{goranci2025fully} show that maintaining a mapping to the dynamic EMD problem that achieves an approximation ratio better than $2$ requires at least $\Omega(n)$ time, even for $1$ dimensional point sets. This highlights the difficulty of dynamically maintaining mappings underlying different proximity measures between point clouds. 

Recently, there has been a growing interest in designing dynamic algorithms for fundamental problems in machine learning, thus contributing towards the grand vision of building a library of efficient data structures for key machine learning primitives. Notable progress has been made on several fronts, including  dynamic algorithms for various clustering objectives such as $k$-center~\cite{chan2018fully,GoranciHLSS21, BateniEFHJMW23, Cruciani2024, BiabaniHM023, Lacki2024, ChanLSW24}, $k$-median/$k$-means~\cite{cohen2019fully, henzinger2020fully, bhattacharya2023fully}, facility location~\cite{GoranciHL18,DBLP:conf/nips/Cohen-AddadHPSS19, bhattacharya2024dynamic}, correlation clustering~\cite{cohen2024dynamic}, as well as dynamic matrix multiplication for structured matrices arising in machine learning applications~\cite{anand2025structural}.

\section{Preliminaries}



\begin{definition}[Chamfer distance]
    Given two point-sets \(A, B\subset \mathbb{R}^d\) with \(\max\{|A|, |B|\}\leq n\), the \textit{Chamfer distance} is defined as $\CH(A, B):=\sum_{a\in A} \min_{b\in B} \dist(a, b)$. For $\alpha \geq 1$, we say that a value $\tilde{\mu}$ is an $\alpha$ approximation to $\CH(A,B)$ if $\tilde{\mu} \leq \CH(A,B) \leq \tilde{\mu} \cdot \alpha$.
\end{definition}

 We will refer to $\min_{b \in B} \dist(a, b)$ as $\CH(a,B)$. 
 We will use the following dynamic nearest-neighbor data structure as a subroutine in our algorithm.

\begin{definition}[Dynamic nearest-neighbor data structure]
\label{def:dynamicNN}

Given \(\alpha > 0\) and a dynamic point set $B \subset \mathbb{R}^d$, $|B| \leq n$, a $(1+\alpha)$-approximate dynamic nearest neighbor data structure with update time and query time $\tau(\alpha)$ with respect to the $\ell_1$ norm is a data structure which can be maintained in $\tau(\alpha)$ update time as \(B\) undergoes insertions and deletions, and when queried for point $a \in \mathbb{R}^d$, it returns a value $\tilde{\mu}_a$ such that $\tilde{\mu}_a \leq \min_{b \in B}\|a-b\|_1 \leq (1+\alpha) \cdot \tilde{\mu}_a $ with $(1-1/\poly(n))$ probability in $\tau(\alpha)$ time. 
    
\end{definition} 

Throughout this paper, we assume that the input points are contained in $[U]^d$ for some $U = \poly(n)$, which is a power of $2$. Furthermore, we assume that during all updates to the input sequence, the \emph{aspect-ratio} of the input points $\max_{a \in A, b\in B}\|a-b\|_1/\min_{a \in A, b \in B}\|a-b\|_1$ is upper bounded by $\phi = \poly(n) = 2^\mathcal{L}$, for some integer \(\mathcal{L}\geq0\).


\begin{lemma}[Dynamic Weighted Sampler]
    \label{lem:sampler}
    There exists a dynamic algorithm that maintains a weighted set of elements $A = \{a_1,\dots,a_n\}$ with corresponding weights $W = \{w(a_1),\dots,w(a_n)\}$ undergoing insertions and deletions. Upon query, the algorithm returns an element of $A$ such that $a_i \in A$ is returned with probability $w(a_i)/\sum_{j \in [n]}w(a_j)$. Both updates and queries are supported in $O(\log n)$ worst-case update and query time.
\end{lemma}

Data structures similar to that of Lemma~\ref{lem:sampler} have appeared before in literature, but for sake of completeness, we provide the proof of the lemma below.

\begin{proof}

The data structure maintains a balanced binary tree of the elements of $A$, where each leaf corresponds to some element. In addition, each node of the binary search tree keeps track of the sum of the weights of the elements corresponding to the leaves in its sub-tree. 

On query, the data-structure draws a random real value $x$ in $[0, \sum_{i = 1}^n w_i]$. Starting from the root of the search tree, it completes a walk to a leaf. When deciding between child nodes $v$ and $u$ (from left to right) with leaf weight sums $w_v, w_u$ in a step of the walk, it chooses $v$ if $x \leq w_v$. Once reaching a leaf, it returns the element assigned to it.

Observe that if the ordering of the leafs from left to right is $a_1 , \dots a_n$, then this process chooses element $a_i$ if $x > \sum_{j < i}w_i$ and $x \leq \sum_{j \leq i} w_i$, that is with probability $w_i / \sum_{j \in [n]} w_j$. As the tree will have depth at most $O(\log n)$, the query takes $O(\log n)$ time. The maintenance of a balanced binary tree of $n$ elements in $O(\log n)$ is folklore.
    
\end{proof}

\subsection{Dynamic Quad-Tree}
\label{sec:dynamic-tree}
 Our algorithm relies on the dynamic quad-tree data structure \cite{de2007efficient}. The quad-tree is constructed as follows. We first choose a random vector described by $z \in [0,U]^d$. We then shift all input points with the vector described by $z$, hence after the shift they will be contained in $[0,2\cdot U]^d$. With a slight overload of notation, we will refer to the shifted points as $A$ and $B$.

Consider a series of $O(\log n)$ grids drawn on the input space, where the $i$-th grid has side length $U \cdot 2^{1-i}$. The quad-tree is then a rooted tree $T$, where each node $v \in T$ is associated with some cell $\mathcal{C}_v$ of these grids with side length $L(v)$.
The root $r$ of $T$ corresponds to the smallest cell of among all the grids which contains all input points. Any node $v$ of the tree $T$ such that $\mathcal{C}_v$ contains more then one point of the input has child nodes in $T$ corresponding to its non-empty sub-cells on the next level of the grid decomposition with side length $L(v)/2$.
The leaves of $T$ corresponds to the largest cells of the decomposition containing a single point of the input.
As the aspect ratio of the input is assumed to be $\phi = \poly(n)$, the tree consists of $O(\log \phi) = \mathcal{L}$ layers.
As each input point may appear once in any of the cells of all these layers, $T$ has at most $O(n \cdot \log n)$ nodes.

\cite{de2007efficient} has shown how to maintain this representation of the input in $O(d \cdot \log n)$ worst-case update time such that all nodes of the tree store the number of input points in their respective cell and the leaves explicitly store the single input point stored in their cell. For our application, we further impose that every cell is aware of the number of input points in its cell from $A$ and $B$ separately.

\section{Dynamic Algorithm}
\label{sec:dynamic}

We start by describing an algorithm that achieves slightly worse guarantees than described in \Cref{thm:main}. Namely, this algorithm will maintain the estimate of the Chamfer distance between two sets with a constant probability. Formally, we will first prove the following result.

\begin{theorem}
    \label{thm:main_const_prob}
    Let \(A,B\) be two sets in space \(\mathbb{R}^d\), with $|A|, |B| \leq n$, \(\alpha > 0\) and \(\epsilon \in (0, 1)\) parameters, and let $\tau(\alpha)$ be the update and query times of a $(1+\alpha)$-approximate dynamic nearest-neighbor datastructure. 
    There exists a dynamic data structure which can be maintained in $O(d \cdot \log n + \log^2 n +  \tau(\Theta(\alpha))$ worst-case update time as $A$ and $B$ undergoes point insertions and deletions and can be queried to return a $(1+\alpha+\epsilon)$-approximation to \(\CH(A,B)\) between $A$ and $B$ w.r.t. \(l_1\)-norm in  $O(\log^2 n \cdot \epsilon^{-2} \max\{\alpha^2,1\} \cdot (d \log^2 n + \tau(\Theta(\alpha))))$ time with $3/4$ probability.
\end{theorem}

In \Cref{subsec:boosting}, we show how to boost the the above result to obtain our main result, i.e. \Cref{thm:main}, which on query returns a $(1+\alpha+\epsilon)$-approximation to $\CH(A,B)$ with probability ($1-1/\poly(n))$. In the following sections, we describe our dynamic algorithm on a high level.

\subsection{Algorithm description}
\label{subsection:algo_desc}

We say that a point $a \in A \cup B$ belongs to a node $v$ of level $i$ of quad-tree $T$ if $a \in \mathcal{C}_v$. We say that point $a \in A$ is matched at $v$ if $v$ is the lowest level level node of $T$ such that both $a$ and any point of $b$ belongs to it (where we assume that the root has the highest level as it corresponds to the largest cell containing all the input points). Note that every point of $A \cup B$ may belong to $\mathcal{L}=O(\log n)$ nodes of $T$, but all points of $A$ are matched at exactly one node of $T$. For each $a \in A$ denote this unique node by $v_a$. 

\subsubsection{Handling an Update}
\label{subsec:updates}

We augment the dynamic tree structure of Section~\ref{sec:dynamic-tree} with the following information being stored at each node \(v \in T\): number of points \(\gamma_A(v)\) from \(A\) belonging to \(v\), number of points \(\gamma_B(v)\) from \(B\) belonging to \(v\), and the number of matched points \(\gamma(v)\) from \(A\) at \(v\). Note that $\gamma_B(v)$ and $\gamma_A(v)$ can be maintained using the algorithm of \cite{de2007efficient}.

We also maintain a dynamic sampler \textsc{Node-Sampler$(v)$} corresponding to each node \(v\) of \(T\), and a global sampler \textsc{Tree-Sampler$(T)$} for the whole \(T\). The sampler \textsc{Node-Sampler$(v)$} is for the set  \(\{u\in T: u \text{ is a child of } v \land \mathcal{C}_u \cap B = \emptyset\}\) w.r.t. weights \(\gamma_A(u)\). The sampler \textsc{Tree-Sampler$(T)$} is for the set of all nodes $v$ of \(T\) with $\gamma(v) > 0$ w.r.t. weights \(w_T(v)=L(v)\cdot \gamma(v)\). 


For sake of simplicity of the presentation, assume that at all times $B \neq \emptyset$. We will now describe how the algorithm updates $\gamma$ values for all $v \in T$ after each update. Assume point $x$ is deleted from or inserted into $A \cup B$. Through iterating along the path starting from the leaf of $T$ containing $x$ and ending at the root $r$, the algorithm finds the path of nodes $v_1,\dots, v_k$ of $T$ whose cell does not contain a point of $B \setminus x$ (ordered from the leaf). Let $v'$ be the ancestor of $v_k$ in $T$. We distinguish between four cases.

{\compactify
\begin{enumerate}
    \item \textbf{Insertion of \(x\) into $A$:} set $\gamma(v') = \gamma(v') + 1$. 
    \item \textbf{Deletion of \(x\) from $A$:} set $\gamma(v') = \gamma(v')-1$. 
    \item \textbf{Insertion of \(x\) into \(B\):} set $\gamma(v') := \gamma(v') - \gamma_A(v_k)$, $\gamma(v_1) = \gamma_A(v_1)$ and $\gamma(v_i) := \gamma_A(v_i) - \gamma_A(v_{i-1})$ for $k \geq i > 1$.
    \item \textbf{Deletion of \(x\) from \(B\):} set $\gamma(v') := \gamma(v') + \sum_{i \in [k]} \gamma(v_i)$ and $\gamma(v_i) = 0$ for $i \in [k]$.
\end{enumerate}
}

For the sake of completeness, the pseudocode for the update operations is provided in \Cref{alg:insert_a}. The correctness of our algorithm in maintaining $\gamma(v), v \in T$ is proven by Claim~\ref{claim:maintain_tree}. 

\begin{algorithm}[H]

\caption{\(\textsc{Update Point}(x)\)}
\label{alg:insert_a}
    \begin{algorithmic}[1]
        \State{ \(v_1 \gets\) leaf containing \(x\)}
        \State{\(\Pi \gets\) path \((v_1,v_2,\dots , r)\) from \(v_1\) to root \(r\) in $T$}
        \State{$v_1, \dots ,v_k \leftarrow v \in \Pi | \mathcal{C}_v \cap (B \setminus x) = \emptyset$}
        \State{$v' \leftarrow$ ancestor of $v_k$ in $T$}
        \If{Insertion into $A$}
            \State{$\gamma(v') \leftarrow \gamma(v') + 1$}
        \EndIf
        \If{Deletion from $A$}
            \State{$\gamma(v') \leftarrow \gamma(v') - 1$}
        \EndIf
        \If{Insertion into $B$}
            \State{$\gamma(v') \leftarrow \gamma(v') - \gamma_A(v_k)$}
            \State{$\gamma(v_1) \leftarrow \gamma_A(v_1)$}
            \State{$\gamma(v_i) \leftarrow \gamma_A(v_i) - \gamma_A(v_{i-1})$ for $k \geq i > 1$}
        \EndIf      
        \If{Deletion from $B$}
            \State{$\gamma(v') \leftarrow \gamma(v') + \sum_{i \in [k]} \gamma(v_i)$}
            \State{$\gamma(v_i) \leftarrow 0$ for $i \in [k]$}
        \EndIf
    \end{algorithmic}
\end{algorithm}

\subsubsection{Answering Queries}
\label{subsec:queries}

The queries are answered using importance sampling based on a sampling process for points in $A$.


\paragraph{Sampling points in \(A\).} To sample a single point from \(A\), we first sample a node \(v\) from \(T\) using \textsc{Tree-Sampler$(T)$}. Then, we sample a child \(u\) of \(v\) using \textsc{Node-Sampler$(v)$}. After this, we recursively call \textsc{Node-Sampler$(u)$}, until we reach a leaf of \(T\), when we finally return the unique point \(a\in A\) contained in it. The details are given in \Cref{alg:sampler} and  \Cref{alg:sampler_find}. 

Claim~\ref{claim:sample:correctness} shows that this sampling process returns point $a \in A$ with probability $L(v_a)/\sum_{v \in T}\gamma(v) \cdot L(v)$, where we recall that $v_a$ stands for the unique cell $a$ is matched in. Claim~\ref{claim:approximation} shows that $L(v_a) \sim \CH(a,B)$ within $\poly(\log n)$ factors.

\begin{algorithm}[H]
\caption{\(\textsc{Sampler}(A)\) procedure}
\label{alg:sampler}
    \begin{algorithmic}[1]
        \State{Sample vertex \(v\) of \(T\) using \textsc{Tree-Sampler$(T)$}}
        \State{Return $\textsc{Sampler-Find}(v)$}
    \end{algorithmic}
\end{algorithm}

\begin{algorithm}[H]
\caption{\(\textsc{Sampler-Find}(v)\) procedure}
\label{alg:sampler_find}
    \begin{algorithmic}[1]
        \If{$v$ is a leaf of quad-tree $T$}
            \State{\textbf{return} $a$ contained in $v$}
        \Else
            \State{Sample child \(u\) of \(v\) using \textsc{Node-Sampler$(v)$}}
            \State{\textsc{Sampler-Find}\((u)\)}
        \EndIf
    \end{algorithmic}
\end{algorithm}

\paragraph{Estimating the Chamfer distance.} Using our sampler, estimating $\CH(A,B)$ turns into a standard application of importance sampling. Function \(\NN(a, B, \alpha)\) refers to any procedure for finding a \((1+\alpha)\)-approximation to the nearest neighbor of \(a\) in set \(B\).

Specifically, we estimate $\CH(A,B)$ through taking $m = 240\cdot \mathcal{L} \cdot \log n \max\{\alpha^2,1\} \cdot \epsilon^{-2} = O(\log^2n \cdot \max\{\alpha^2,1\} \cdot \epsilon^{-2})$ samples $S$ from $A$ through the after-mentioned sampling procedure. For each $a \in S$, we then query the nearest neighbor data structure to generate a $(1+\alpha/4)$-approximation to $\CH(a,B)$. Refer to these values as $\NN(a,B,\alpha/4)$ for $a \in S$. 

We assign a weight of $\NN(a,B,\alpha/4) \cdot \sum_{v \in T}\gamma(v) \cdot L(v)/L(v_a)$ to each $a \in S$. This implies that the weight of each sample is a $(1+\alpha/4)$-approximation of $\CH(A,B)$ in expectation. We then finally return the average of these weights (shifted by $1/(1+\epsilon/2)$ to fit our definition of approximation). The details are given in \Cref{alg:chamfer}.

\begin{algorithm}[H]
\caption{\(\textsc{Query-Chamfer}(A, B, \epsilon, \alpha)\) procedure}
\label{alg:main}
\label{alg:chamfer}
    \begin{algorithmic}[1]
    \State{$\tilde{\mu} = 0$}
    \State{$m = \frac{120 \cdot \mathcal{L} \cdot \log n \max\{\alpha^2,1\}}{\epsilon^2}$} \Comment{$m = O(\frac{\log^2n \cdot \max\{\alpha^2,1\}}{\epsilon^2})$}
    \For{$i = 1$ to $m$}
        \State $a \leftarrow \textsc{Sampler}(A)$
        \State $x_a \gets \NN(a,B,\alpha/4) \cdot \frac{\sum_{v \in T}\gamma(v) \cdot L(v)}{L(v_a)}$
        \State $\tilde{\mu} = \tilde{\mu} + x_a$ 
    \EndFor
    \State \ Return $\frac{\tilde{\mu}}{m \cdot (1+\epsilon/2)}$
    \end{algorithmic}
\end{algorithm}


\subsection{Correctness Analysis}

This section is devoted to an overview of the proof of Lemma~\ref{lemma:correctness}, which establishes the correctness of our algorithm.

\begin{lemma}
\label{lemma:correctness}
    On query, the algorithm returns a $(1+\epsilon+\alpha)$-approximation to $\CH(A,B)$ with $7/8$ probability.
\end{lemma}



We start with examining the variables $\gamma(v)$, corresponding to the number of points from $A$ matched at node $v$.

\begin{claim}
\label{claim:maintain_tree}
    The value of $\gamma(v)$ for all \(v\in T\) is maintained correctly after an update.
\end{claim}

\begin{proof}

    When an update occurs to $a \in A$, the only $\gamma$ value changes is $\gamma(v_a)$. The algorithm greedily finds $v_a$ and updates $\gamma(v_a)$ accordingly.

    When an update occurs to $b \in B$, the algorithm finds all nodes of $T$ that contain only $b$ from $B$, $v_1,\dots,  v_k$ (ordered from the leaf upwards) and $v'$ the node with the smallest cell containing $b$ and an other point of $B$. Observe that only $v_1, \dots, v_k$ and $v'$ may have its $\gamma$ value updated.

    In the case of a deletion, all points of $A$ matched in $v_1, \dots,  v_k$ should be matched in $v'$ after an update, and the algorithm updates $\gamma$ values accordingly. In case of an insertions, every point of $A$ in $\mathcal{C}_{v'}$ should be matched in the smallest cell among $\mathcal{C}_{v_1}, \dots, \mathcal{C}_{v_k}$ they are contained in (if there is such a cell). The algorithm similarly updates $\gamma$ values accordingly.
\end{proof}

The following claim allows us to estimate $\CH(a,B)$ with value $L(v_a)$. 
Our proof is similar to a lemma of \cite{bakshi2023near}, however our lower bound on $L(v_a)$ with respect to $\CH(a,B)$ is slightly weaker due to the limitations of the dynamic model.

\begin{claim}
\label{claim:approximation}

The following statements hold with respect to  the random shift of the quad-tree (see \Cref{sec:dynamic-tree}). 


{\compactify
\begin{enumerate}
    \item[(i)] \label{item1} With $1-1/\poly(n)$ probability $L(v_a) \geq \frac{\CH(a,B)}{\log n \cdot 3}$ holds for all $a \in A$.
    \item[(ii)] \label{item2} $\E[L(v_a)] \leq 2 \cdot \mathcal{L} \cdot \CH(a,B))$ holds for all $a \in A$, where \(\mathcal{L}\) denotes the height of \(T\).
\end{enumerate}
}

\end{claim}

\begin{proof}

We will first prove some useful properties of the underlying quad-tree datastructure, which are based on the random shift of the points introduced at initialization. Our proof is analogous to a similar separation lemma of \cite{bakshi2023near}, and we include it here for sake of completeness.

Namely, let $x,y \in A \cup B$ and $v$ be some node of $T$. Then, with respect to the random shift defined by $z \in [0,U]^d$, we have:

\begin{enumerate}
    \item \(\Pr [y \notin \mathcal{C}_v | x \in \mathcal{C}_v]\leq \frac{\|x-y\|_1}{L(v)}\)
    \item \(\Pr [y \in \mathcal{C}_v | x \in \mathcal{C}_v]\leq \exp(-\frac{\|x-y\|_1}{L(v)})\)
\end{enumerate}

The random shift introduced at the initialization of our algorithm can be described as follows: the algorithms draws a uniform random point $z$ with coordinates $z_1, \dots z_d$ in $[U]^d$. Then the algorithm defines sets $A: = \{a + z |a \in A\}$ and $B = \{b + z| b \in B\}$.

Fix some cell $\mathcal{C}_v = [v_1, v_1 + L(v)] \times \dots \times [v_d, v_d + L(v)]$ (we disregard the case where a shifted point falls exactly on the grid, as it occurs with $0$ probability). If point $x \in A \cup B$ with coordinates $x_1, \dots, x_d$ falls in cell $\mathcal{C}_v$, we know that $z_i \in [x_i-v_i, x_i-v_i + L(v)]$ for all $i \in [d]$. If we condition on this event, then we know that $z_i$ is uniformly distributed on this interval. 

Fix some $y \in A \cup B$ with coordinates $y_1, \dots, y_d$. Conditioning on the event that $x \in \mathcal{C}_v$, the event $y \notin \mathcal{C}_v$ is for all $i \in [d]$ if $z_i \in [x_i-v_i, x_i - v_i + L(v)], \notin [y_i-v_i, y_i-v_i + L(v)]$ that is $|x_i-y_i|> L(v)$ or $z_i \in [x_i - v_i, y_i - v_i]$ or $z_i \in [y_i-v_i + L(v), x_i-v_i+L(v)]$ (depending on which of $x_i,y_i$ is larger) which happens with probability at most $|x_i - y_i|/L(v)$. Hence, by union bounding over these events:

$$\Pr[y \notin \mathcal{C}_v | x \in \mathcal{C}_v] \leq \sum_{i = 1}^d\frac{|x_i-y_i|}{L(v)} = \frac{\|x-y\|_1}{L(v)}$$

Now observe that the above argument holds for all coordinates of $z$ independently. This implies:

$$\Pr[y \in \mathcal{C}_v| x \in \mathcal{C}_v] \leq \prod_{i = 1}^d (1 - \frac{|x_i-y_i|}{L(v)}) \leq \exp(-\frac{\|x-y\|_1}{L(v)})$$

We are now ready to prove the claim. Fix some $a \in A$. If $L(v_a) < \CH(a,B) \cdot 1/(4\cdot \log n)$, that implies that for some $U/2^{i_0} \leq \CH(a,B)/(3 \cdot \log n)$ there exists a $b \in B$ such that both \(a\) and \(b\) belong to the same cell in the tree $T$ with side length $U/2^{i_0}$. Let $\mathcal{C}_a^{i_0}, \dots ,\mathcal{C}_a^k$ be the cells of $T$ with side length at most $U/2^{i_0}$ that $a$ belongs to. Recall that this means $k \leq \mathcal{L}$. By property (ii), we know that for all $b \in B$:

\begin{align*}
\sum_{i \in \{i_0, \dots, k\}}\Pr[b \in \mathcal{C}^i_a | a \in \mathcal{C}^i_a] & \leq  \sum_{\{i_0, \dots, k\}}\exp\left(\frac{\|a - b\|_1}{2^i}\right) \nonumber \\
& \leq \mathcal{L}  \cdot \exp(\frac{\|a-b\|_1}{2^i})\nonumber \\
& \leq  \mathcal{L}  \cdot\exp\left(\frac{\CH(a,B)}{2^i}\right) \nonumber \\
& \leq  \mathcal{L}  \cdot \exp(-3\cdot \log n) \nonumber \\
& \leq  1/n^{3} \nonumber
\end{align*}


By union bounding over all $b \in B$ and $a \in A$, and observing that these inequalities hold regardless which cells $a$ falls on in specific levels of the tree, we get that the first item holds with high $1-\poly(n)$ probability for all $a \in A$. 

To prove point (ii), fix some $a \in A$ and $b = \argmin_{b \in B} \|a-b\|_1$. Let $\mathcal{C}_{a}^{i_0}, \dots, \mathcal{C}_a^0$ be the cells of $T$ with size lengths at least $\CH(a,B)$ containing $a$ where $\mathcal{C}_a^i$ has side length $U/2^i$. If $L(v_a) \geq U/2^{i-1}$ then $b \notin \mathcal{C}_a^i$. Hence, by property (i) we have that $\Pr[L(v_a) \geq U/2^{i-1}] \leq \Pr[b \notin \mathcal{C}_a^i | a \in \mathcal{C}_a^i] \leq 2^{i} \cdot \|a-b\|_1/U$. Note that $L(v_a) \leq L(\mathcal{C}_a^0)$ by definition. Hence, summing over the $\mathcal{L}$ levels of the tree we get:

\begin{align*}
\E[L(v_a)] & \leq \sum_{\{i_0, \dots, 0\}}\Pr[L(v_a) \geq \frac{U}{2^{i-1}}] \cdot L(\mathcal{C}_a^i) \nonumber \\
& \leq \sum_{\{i_0, \dots, 0\}} \Pr[b \notin \mathcal{C}_a^i | a \in \mathcal{C}_i^i] \cdot \frac{U}{2^{i-1}} \nonumber \\
& \leq \sum_{\{i_0, \dots, 0\}} \frac{2^i \cdot \|a-b\|_1}{U}\cdot \frac{U}{2^{i-1}} \nonumber \\
& \leq \sum_{\{i_0, \dots, 0\}} 2 \cdot \|a - b\|_1\nonumber \\
& \leq 2 \cdot \mathcal{L} \cdot \CH(a,B)\nonumber
\end{align*}

Note that the second inequality holds as it does not matter which cells $a$ falls on different levels of the tree for the set of proceeding inequalities.

\end{proof}

From Claim~\ref{claim:approximation} we have that $L(v_a) \sim \CH(a,B)$. Hence, the following claim shows that the sampling process of the algorithm samples $a \in A$ with probability roughly proportional to its contribution to $\CH(A,B)$. 

\begin{claim}
\label{claim:sample:correctness}
The algorithm samples $a \in A$ with probability $L(v_a)/\sum_{v \in T}L(v) \cdot \gamma(v)$.
\end{claim}

\begin{proof}
        Fix a point \(a\in A\), and let \(X\) be the event when \(a\) is sampled in \Cref{alg:sampler}. Let \(v_a, v_2, ..., v_l\) be the path on tree starting at $v_a$ ending at the leaf \(v_l\) containing $a$ (hence $\gamma_A(v_l) = 1$). First, denote by \(Y\) the event when \textsc{Tree-Sampler$(T)$} returns node \(v_a\), and by \(X_i\) the event that \textsc{Node-Sampler$(v)$e} in line 4 of \Cref{alg:sampler_find} choose the node \(v_i\) on level \(i\) of \(T\), for \(1<i\leq l\). By definition, we have that \(\Pr[Y]=\frac{L(v_a)\cdot \gamma(v_a)}{\sum_{v\in T} L(v)\cdot \gamma(v)}\). Further, 
    \[\Pr[X_2|Y]=\frac{w(v_2)}{\sum_{u: u \in \mathbb{C}(v_1)} w(u)}=\frac{\gamma_A(v_2)}{\gamma(v_a)}, \]
    by definition of \(w(\cdot)\).
    Similarly, for \(3\leq i\leq l\)
    \[\Pr[X_i|X_{i-1}]=\frac{w(v_i)}{\sum_{u: u \in \mathbb{C}(v_{i-1})} w(u)}=\frac{\gamma_A(v_i)}{\gamma_A(v_{i-1})}\]

    Finally, we have that 

\begin{align*}
\Pr[X] & = \Pr[Y\cap X_2\cap ...\cap X_l] \nonumber \\
& = \Pr[Y] \cdot \frac{\gamma_A(v_2)}{\gamma(v)} \cdot \prod_{i=3}^l \frac{\gamma_A(v_i)}{\gamma_A(v_{i-1})} \nonumber \\
& = \Pr[Y]\cdot \frac{\gamma_A(v_l)}{\gamma(v_a)}\nonumber \\
& = \frac{L(v_a)}{\sum_{v \in T}L(v) \cdot \gamma(v)}. \nonumber
\end{align*}

\end{proof}




    


The remaining proofs follow the standard analysis for importance sampling. Our main goal is to bound the variance of a single random variable defined by the weight assigned to each of the $m$ samples of $A$ the algorithm draws on query. 

\begin{claim}
\label{claim:variance}
Let $X$ stand for the random variable defined by $NN(a,B,\alpha/4) \cdot \sum_{v \in V}\gamma(v) \cdot L(v)/L(v_a)$ when the algorithm samples point $a \in A$. Then, with high ($1-1/\poly(n)$) probability $\Var[X] \leq CH(A,B)^2 \cdot \log n \cdot \mathcal{L} \cdot 12 \alpha^2$.
\end{claim}

\begin{proof}

First, note that when we decide to query the distance of some $a \in A$ from $B$, we use a $(\alpha+\epsilon/4)$-approximate nearest neighbor datastructure, which might be randomized. Hence, instead of returning $\CH(a,B)$, we obtain some $Z_a \cdot \CH(a,B)$, where $Z_a$ is a random variable taking real values in $[1, 1+\alpha /4+\epsilon/4]$. Note that this implies that $\E[Z_a^2] \leq ( 1+\alpha /4+\epsilon/4)^2\leq 2\alpha^2$. Also, note that the randomization of $Z_a | a \in A$ is independent from the randomization of the rest of the algorithm (i.e. the randomness used for sampling elements of $A$ and that of $\mathcal{H}$). Furthermore, the randomness we use to sample $a \in A$ is independent of the randomness of the quad-tree (that is of $L(v_a)$ and $\gamma(v)$ values).




\begin{align*}
\Var[X] & \leq   \E[X^2] \nonumber \\
& = \sum_{a \in A}\E\left[\frac{L(v_a)}{\sum_{v \in T} L(v) \cdot \gamma(v)}\frac{(\sum_{v \in T}\gamma(v) \cdot L(v))^2}{L(v_a)^2}\right] \cdot \E[Z_a^2] \cdot \CH(a,B)^2 \nonumber \\
& =  \sum_{a \in A}\E\left[\frac{\sum_{v \in T} \gamma(v) \cdot L(v) \cdot \CH(a,B)}{L(v_a)}\right] \cdot \E[Z_a^2] \cdot \CH(a,B)\nonumber \\
& \leq  \E[\sum_{v \in T}\gamma(v) \cdot L(v) ] \cdot \sum_{a \in A} \log n \cdot \CH(a,B) \cdot 6 \alpha^2\nonumber \\
& \leq  \log^2 n \cdot \mathcal{L} \cdot \CH(A,B)^2  \cdot 12\alpha^2\nonumber
\end{align*}

The second inequality holds due to Claim~\ref{claim:approximation}, part (i), with high ($1-1/\poly(n)$) probability. The last inequality follows from the same claim, part (ii).
    
\end{proof}

We are now ready to finally prove Lemma~\ref{lemma:correctness}, which is a standard application of importance sampling. Given random variable $X$ with expectation $\E[X]$ and variance $\E[X]^2 \cdot \phi$, the average of $O(\phi/\epsilon^2)$ i.i.d. samples of $X$ is a $(1+\epsilon)$-approximation to $\E[X]$ with $>1/2$ probability by Chebyshev's inequality.

In our case, $\E[X]$ is $(1+\alpha/4)$-approximate to $\CH(A,B)$ and $\phi = O(\alpha^2 \epsilon^{-2}\log^2 n )$ by Claim~\ref{claim:variance}. Hence, we conclude the main lemma of this section \Cref{lemma:correctness}. 


\begin{proof}[Proof of \Cref{lemma:correctness}]

On query, the algorithm returns the value of random variable $\tilde{\mu}/(m \cdot (1+\epsilon/2))$. First, in order for this to be a $(1+\alpha+\epsilon)$-approximation to $\CH(A,B)$, by definition it must hold that

\begin{equation}
\label{eq:estimate}
\frac{\tilde{\mu}}{m \cdot (1+\epsilon/2)} \leq \CH(A,B)\leq \frac{\tilde{\mu}}{m \cdot (1+\epsilon/2)} \cdot (1+\alpha+\epsilon).
\end{equation}

Let $X$ be a random variable described by Claim~\ref{claim:variance}, that is a random variable taking value $\NN(a,B,\alpha/4) \cdot \sum_{v \in T}\gamma(v) \cdot L(v)/L(v_a)$ with probability $L(v_a)/\sum_{v \in T} \gamma(v) \cdot L(v)$. By the definition of the nearest neighbor oracle we have that $\CH(A,B)/(1+\alpha/4) \leq\E[X] \leq \CH(A,B)$.

Since $\tilde{\mu}/\alpha$ is the average of $m$ i.i.d. copies of $X$, then $\Var[\tilde{\mu}/m] \leq \Var[X]/m$ and hence 
\begin{equation}
\label{eq:var}
\Var[\tilde{\mu}/m]\leq \Var[X] \cdot \frac{\epsilon^2 \cdot \log^2 n}{240 \cdot \alpha^2}\leq \frac{\CH(A,B)^2 \cdot \epsilon^2}{20 \cdot \max\{\alpha^2, 1\}}
\end{equation}
with high probability by Claim~\ref{claim:variance}. Now, by a simple application of Chebyshev's inequality we have

\begin{align*}
 \Pr\left[\left\lvert\frac{\tilde{\mu}}{m} - \E\left[\frac{\tilde{\mu}}{m}\right]\right\rvert > 4 \cdot \sqrt{\Var\left[{\frac{\tilde{\mu}}{m}}\right]}\right] & \leq \frac{1}{16} &  \nonumber. 
 \end{align*}
Since by linearity of expectation it holds that \(\CH(A, B)/(1+\alpha/4)\leq \mathbb{E}[\frac{\tilde{\mu}}{m}]\leq \CH(A, B)\), we have 
 \begin{align*}
\Pr\left[ \frac{\tilde{\mu}}{m} > \CH(A,B) + 4 \cdot \sqrt{\Var\left[\frac{\tilde{\mu}}{m}\right]}\right] + \Pr\left[  \frac{\tilde{\mu}}{m} < \frac{\CH(A,B)}{1+\alpha/4} - 4 \cdot \sqrt{\Var\left[\frac{\tilde{\mu}}{m}\right]}\right]  & \leq \frac{1}{16}.
\end{align*}
Plugging in \ref{eq:var}, we obtain 
\begin{align*}
\Pr\left[ \frac{\tilde{\mu}}{m} > \CH(A,B) +  \frac{\epsilon\cdot \CH(A, B)}{\sqrt{5} \cdot \max\{\alpha,1\}}\right] + \Pr\left[  \frac{\tilde{\mu}}{m} < \frac{\CH(A,B)}{1+\alpha/4} - \frac{\epsilon\cdot \CH(A, B)}{\sqrt{5} \cdot \max\{\alpha,1\}}\right]  & \leq \frac{1}{16}. 
\end{align*}


For all $\alpha >0$ and $0<\epsilon <1$ this implies that:

\begin{align*}
 \Pr\left[ \frac{\tilde{\mu}}{m \cdot (1+\epsilon/2)} > \CH(A,B) \right] + \Pr\left[ \frac{\tilde{\mu}}{m \cdot (1+\epsilon/2)} < \CH(A,B)\cdot (1+\alpha+\epsilon)\right]   & \leq \frac{1}{16}. \nonumber
\end{align*}

Therefore, the output of the query is a $(1+\alpha+\epsilon)$-approximation to $\CH(A,B)$ with $15/16>3/4$ probability.
    
\end{proof}

\subsection{Running time}

\begin{lemma}
\label{lemma:running_time}
    The algorithm of Section~\ref{subsection:algo_desc} has $O(d \cdot \log n + \log^2 n + \tau(\Theta(\alpha))$ and $O(\log^2 n  \cdot \epsilon^{-2} \max \{1,\alpha^2\} \cdot (d\log^2 n +\tau(\Theta(\alpha))))$ worst-case update and query times, respectively. 
\end{lemma}

\begin{proof}
    Note that updating the quad tree itself (and the corresponding $\gamma_A, \gamma_B$ values) takes $O(d \cdot \log n)$ time using algorithms from literature \cite{de2007efficient}. Observe that insertions and deletions both to $A$ and $B$ require the algorithm to identify the set of vertices of $T$ the affected point falls in and simply to adjust the $\gamma$ value on a subset of these nodes. Hence, this requires at most $O(d \cdot \mathcal{L}) = O(d \cdot \log n)$ time. 

    The algorithm also needs to update its internal global \textsc{Tree-Sampler$(T)$} and \textsc{Node-Sampler}s for all $v \in T$. Observe that the total number of updates these samplers undergo is proportional to the number of nodes of $T$ which change their $\gamma_v(A)$ or $\gamma_v$ values, that is $O(\log n)$. By Lemma~\ref{lem:sampler} this takes $O(\log^2 n)$ time as each sampler contains at most $|A| \leq n$ elements. In addition, the algorithm needs to maintain its dynamic nearest-neighbor datastructure.

    On query the algorithm needs to sample $m = O(\log ^2 n \cdot \max\{1,\alpha^2\} \cdot \epsilon^{-2})$ samples, for each of which it queries the nearest neighbor datastructure. Each query requires first a call to \textsc{Tree-Sampler$(T)$}, then a walk from the sampled node to a leaf at each step of which a \textsc{Node-Sampler} is queried. Hence, this takes $O(d\log^2 n \cdot m) = O(d\log^4 n\cdot \max\{1,\alpha^2\} \cdot \epsilon^{-2})$ time.
\end{proof}

\subsection{Boosting for a High-Probability Guarantee After Every Update}
\label{subsec:boosting}

We show to derive \Cref{thm:main} from \Cref{thm:main_const_prob}. To this end, while maintaining the internal data structures of \Cref{thm:main_const_prob}, on query output the median result of $O(\log n)$ queries to the baseline algorithm. This will result in the same update time, but an $O(\log n)$ blowup in query time.

In terms of correctness, consider the $\log n$ outputs the algorithm produces. Each of them is an $(1+\alpha+\epsilon)$-approximation to $\CH(A,B)$ with $3/4>1/2$ probability. By a standard application of Chernoff bound with high $(1-1/\poly(n))$ probability more than half of these query results will be $(1+\alpha+\epsilon)$-approximation, hence so is their median value.

\section{Experiments}
\label{sec:exp}

\begin{figure}[H]
    \centering
    \begin{subfigure}{0.95\textwidth}
        \centering
    \includegraphics[width=\linewidth]{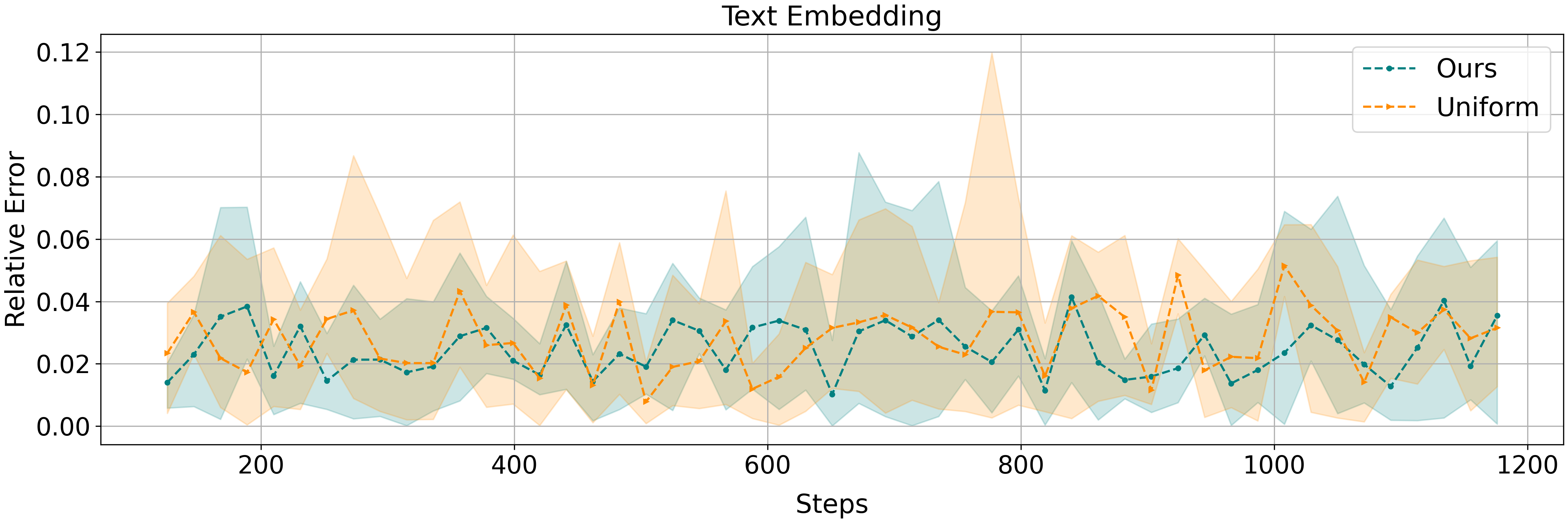}
    \end{subfigure}
    \begin{subfigure}{0.95\textwidth}
        \centering
    \includegraphics[width=\linewidth]{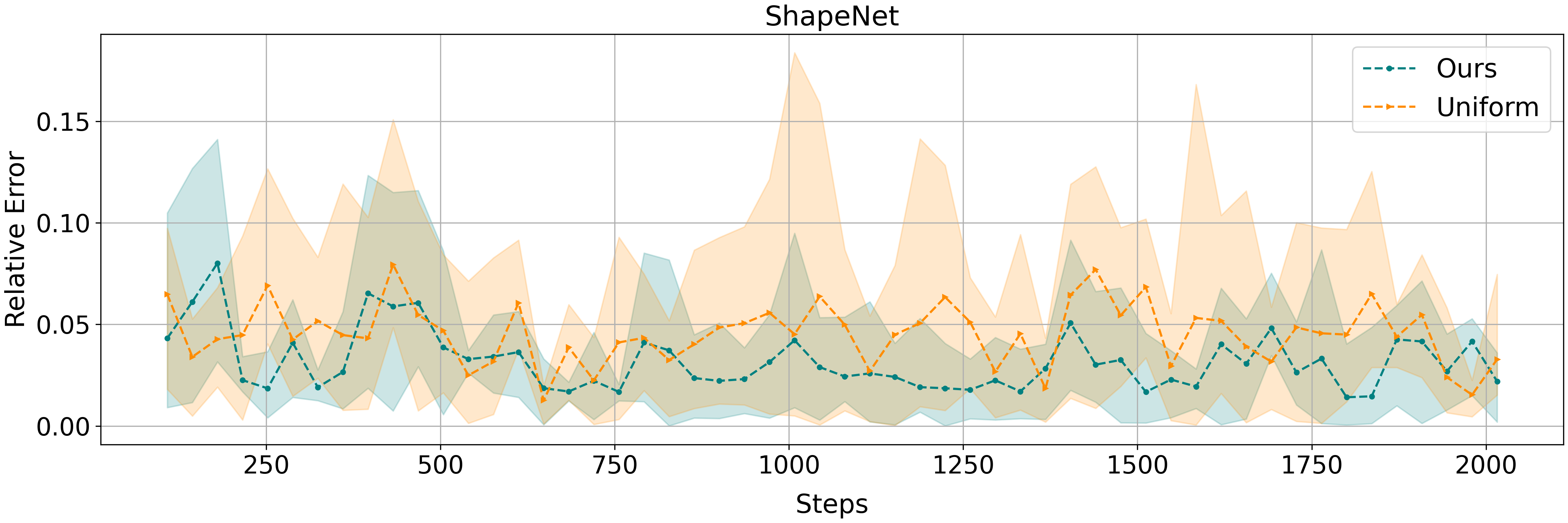}
    \end{subfigure}
    \begin{subfigure}{0.95\textwidth}
        \centering
    \includegraphics[width=\linewidth]{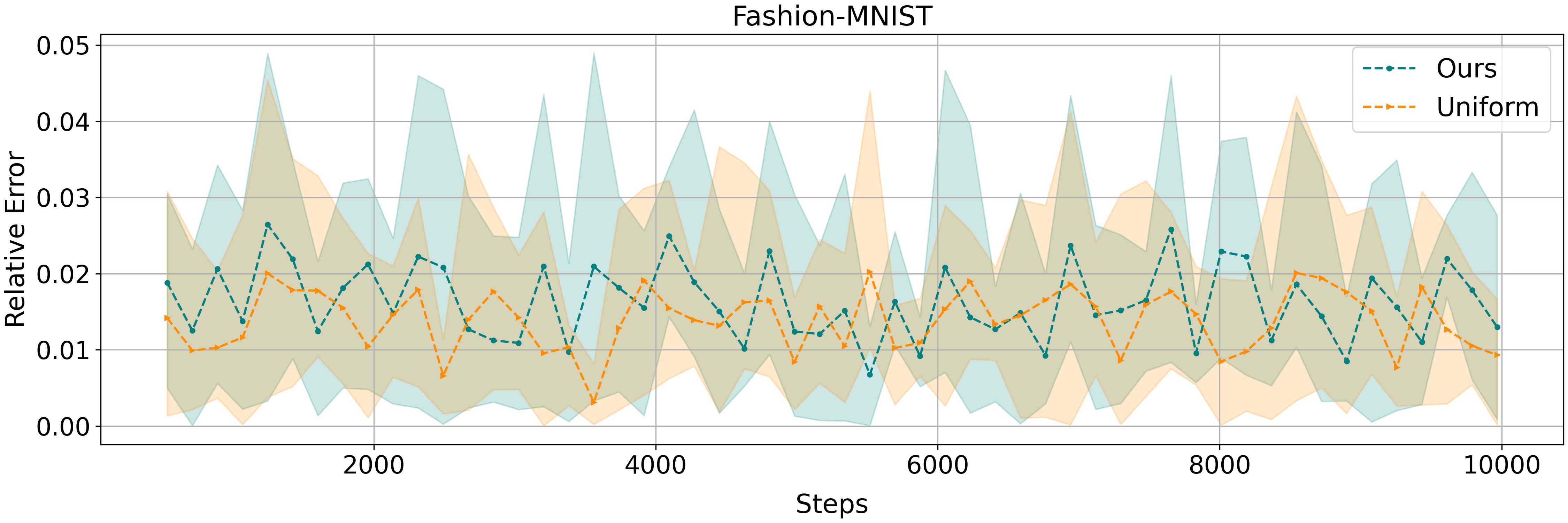}
    \end{subfigure}
    \begin{subfigure}{0.95\textwidth}
        \centering
    \includegraphics[width=\linewidth]{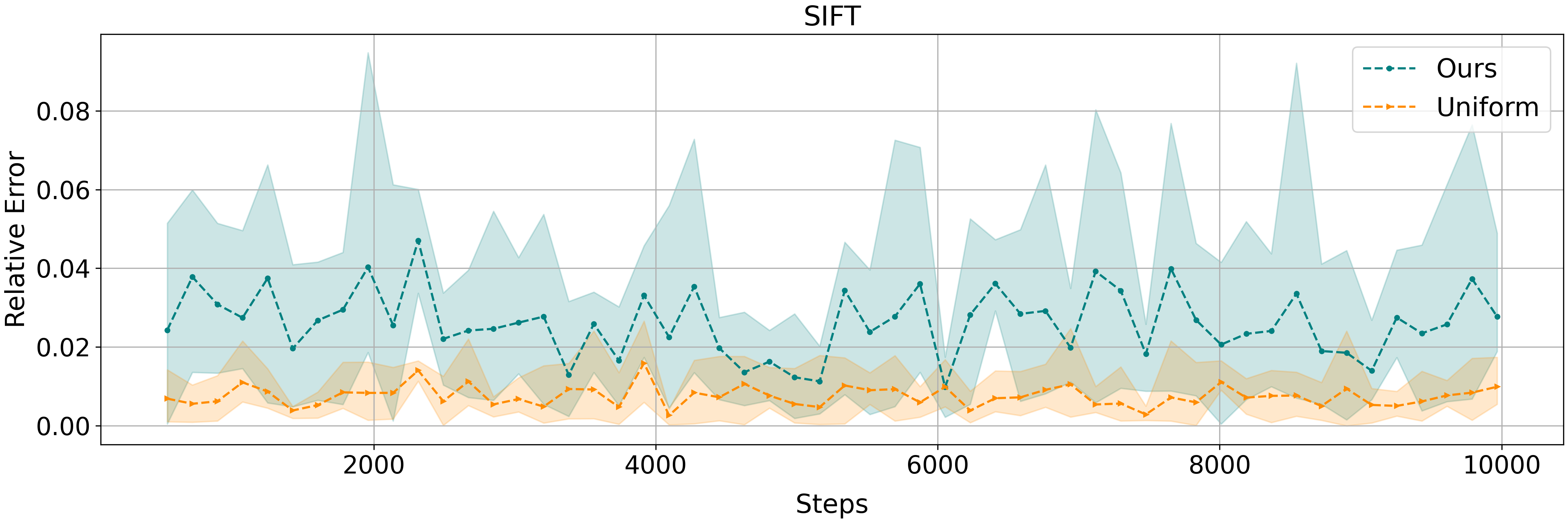}
    \end{subfigure}
    \caption{Relative error curves for datasets \emph{without} outliers.
    These experiments are independently run for $5$ times, and we report the average (the dot), max and min value (the shaded area) after every $\frac{w}{5}$ to $\frac{w}{3}$ updates (depending on the dataset) where $w$ is the window size.
    }
    \label{fig:exp:err}
\end{figure}

\begin{figure}[H]
    \centering
    \begin{subfigure}{0.95\textwidth}
        \centering
    \includegraphics[width=\linewidth]{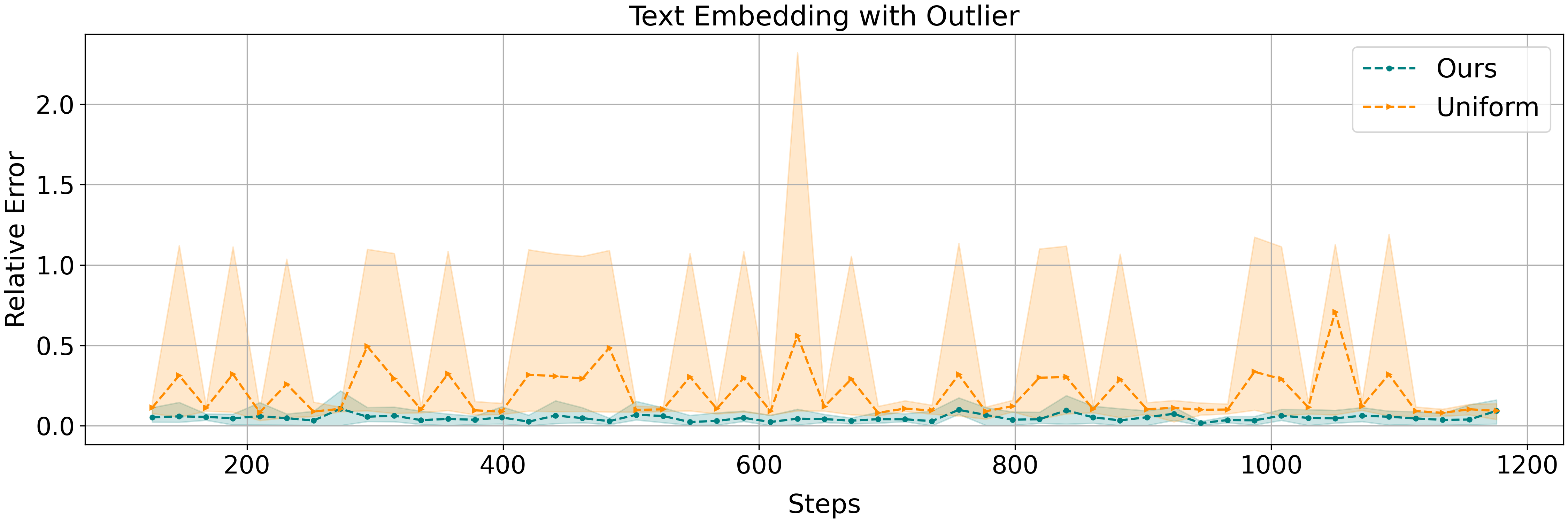}
    \end{subfigure}
    \begin{subfigure}{0.95\textwidth}
        \centering
    \includegraphics[width=\linewidth]{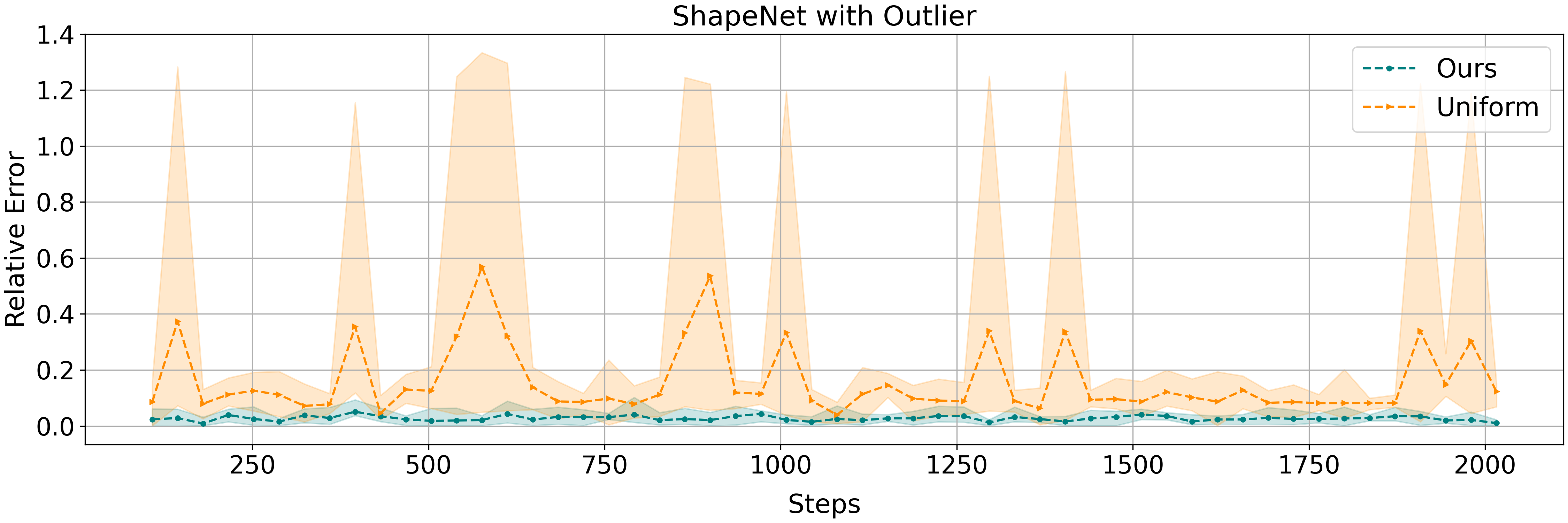}
    \end{subfigure}
    \begin{subfigure}{0.95\textwidth}
        \centering
    \includegraphics[width=\linewidth]{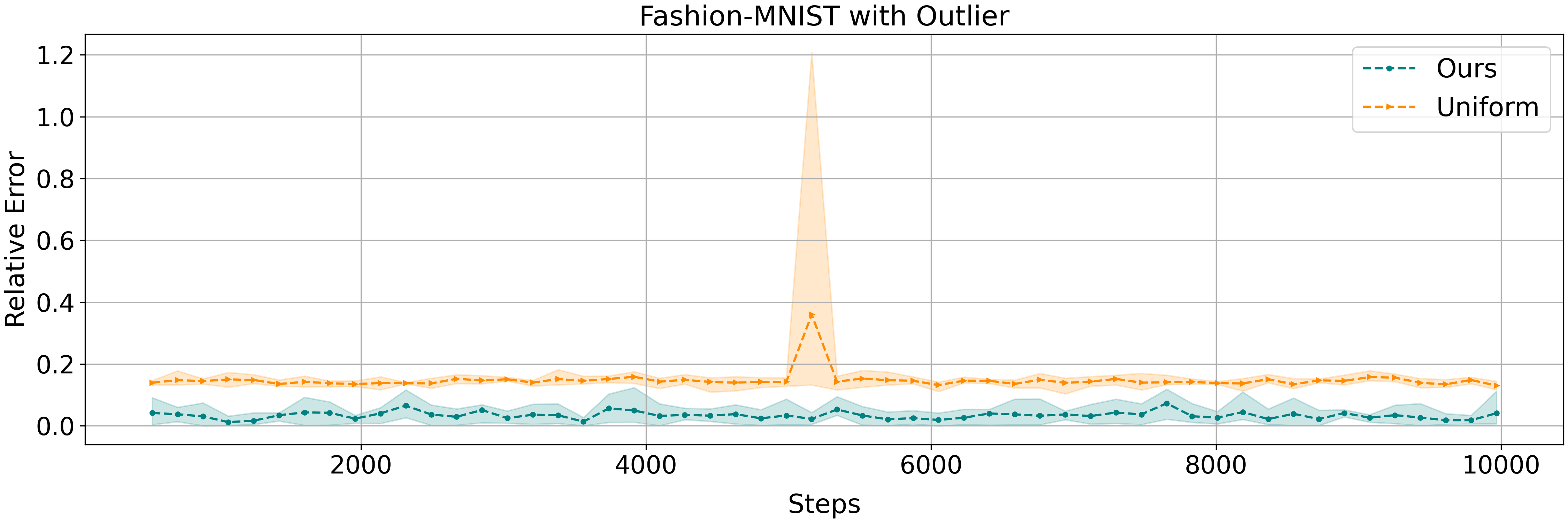}
    \end{subfigure}
    \begin{subfigure}{0.95\textwidth}
        \centering
    \includegraphics[width=\linewidth]{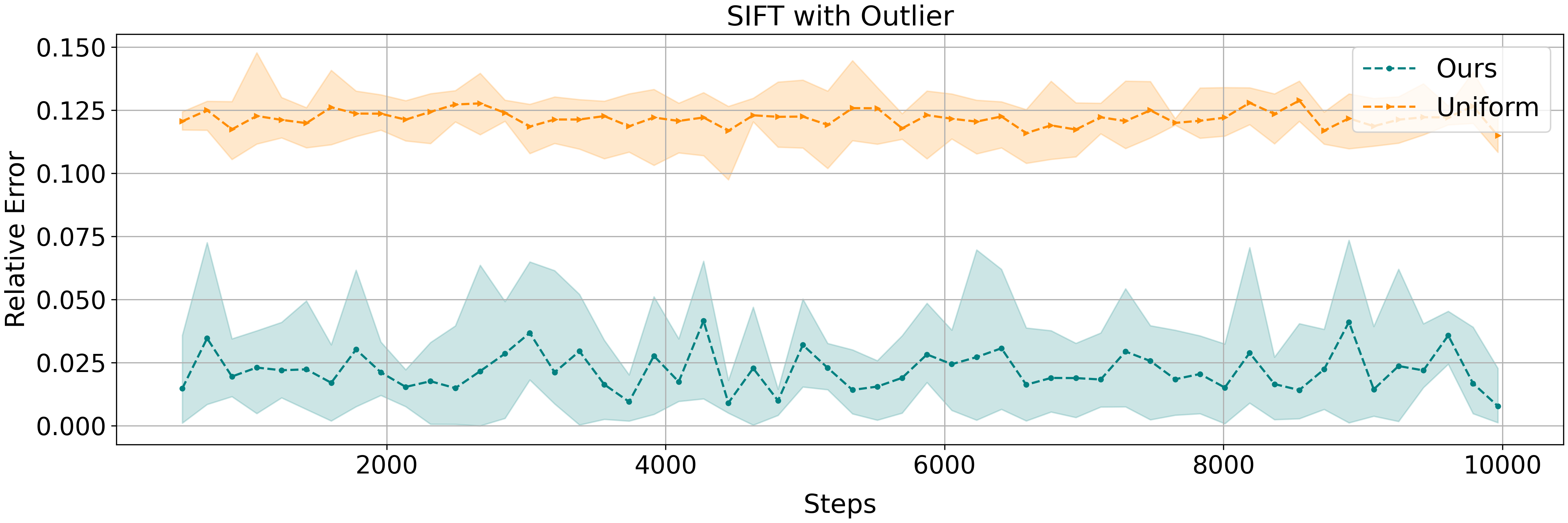}
    \end{subfigure}
    \caption{Relative error curves for datasets \emph{with} outliers, with the same setup as in \Cref{fig:exp:err}.}
    \label{fig:exp:err_outlier}
\end{figure}

\begin{table}[H]
\centering
\caption{Specifications of datasets and experiment parameters.}
\begin{small}
\begin{tabular}{llllll}
\toprule
dataset  & dimension $d$ & $\left|A\right|$ & $\left|B\right|$ & window size & sample size\\
\midrule
Text Embedding    & 300   & \textasciitilde 1.9k  & \textasciitilde 1.2k & 100 & 150\\
ShapeNet          & 3      & \textasciitilde 2k   & \textasciitilde 2k  & 100  & 150\\
Fashion-MNIST     & 784   & 60k & 10k & 500 & 200\\
SIFT & 128  & 1000k & 10k & 500 & 300\\
\bottomrule
\label{tab:dataset}
\end{tabular}
\end{small}
\vspace{-0.2in}
\end{table}

\begin{figure}[H]
        \centering
    \includegraphics[width=0.95\textwidth]{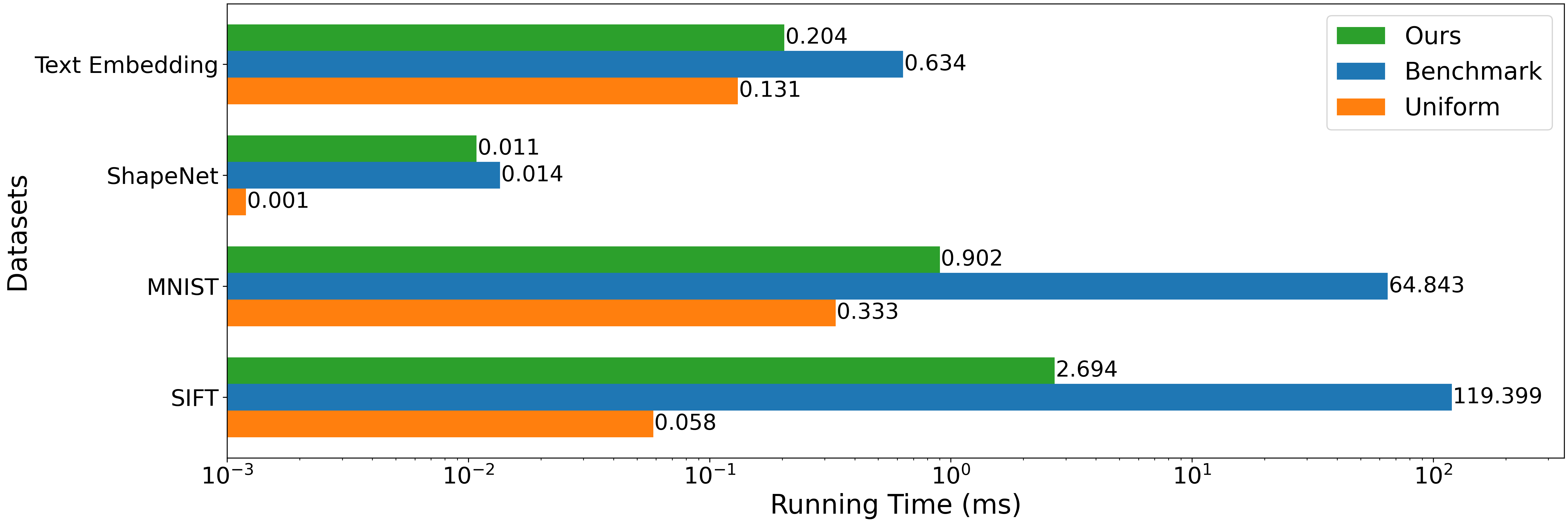}
    \caption{Average running time per window update for all algorithms on datasets \emph{without} outliers.}
    \label{fig:exp:run_time}
\end{figure}

We implement our dynamic algorithm and validate its performance on various datasets.
Observe that handling update of $A$ is straightforward (by simply querying an NN oracle on $B$),
whereas allowing updates on $B$ is more challenging.
Hence, to simplify the exposition (but still keeping the key challenge),
we focus on the default setting where $B$ is dynamic and $A$ is a static set.
For completeness, we also include the case that both $A$ and $B$ are dynamic,
and the results are similar as in the default setting.


\paragraph{Baselines.}
Our first baseline is a naive exact algorithm, which we call ``Benchmark'',
where for each update of $B$, it re-computes $\dist(a, B)$ in $O(d|B|)$ time for each $a$ whose current $\dist(a, B)$ value may be affected by the update.
This serves as a benchmark for the approximation ratio.
The second baseline, called ``Uniform'', replaces our important sampling with a uniform sampling (while keeping the other steps the same).

\paragraph{Datasets and Experiment Setup.}
We employ four real datasets covering both high and low dimensions in the experiment:
Text Embedding~\cite{KusnerSKW15}, ShapeNet~\cite{DBLP:journals/corr/ChangFGHHLSSSSX15}, Fashion-MNIST~\cite{mnist_dataset}, and SIFT~\cite{JegouDS11}.
Each dataset consists of a larger set which we use as $A$, and a smaller set which we use as $B$. The ShapeNet dataset consists of 3D point clouds and is widely used to measure the similarity between different shapes.
The Fashion-MNIST and SIFT datasets were also known to be used as benchmarks for approximate nearest neighbor search.
Similar choice of using ANN benchmark datasets was also made in a previous work, to evaluate the performance of static algorithm for Chamfer distance~\cite{bakshi2023near}.
Furthermore, to evaluate the robustness of the algorithms,
we inject an outlier point into each dataset:
we compute the geometric mean $c := \frac{1}{\left|A\right|}\sum_{a\in A}a$, pick an arbitrary $a^* \in A$,
and generate an outlier as $\tilde{a} = 0.1 \cdot  |A| \cdot (a^* - c) + c$.
Intuitively, this ``moves'' $a^*$ along the direction of $a^* - c$ by a large distance.
Finally, since these datasets do not contain information of dynamic update,
we employ a sliding window (on $B$) to simulate the insertions and deletions.
The detailed specification of the dataset and the experiment parameters can be found in \Cref{tab:dataset}.



\paragraph{Implementation Details.}
Recall that our algorithm (and the Uniform baseline) consists of a sampling step
and then a second step to build the estimator that makes use of a nearest neighbor query structure.
Since the window size for each dataset is relatively small,
we choose to use the \emph{exact} nearest neighbor algorithm, which does not introduce additional errors and allows for a more accurate evaluation.  
For the 3D ShapeNet dataset, we implement nearest neighbor queries with KD-trees, which can efficiently perform exact search in low-dimensional spaces.
All algorithms are implemented in C++ and compiled with Apple Clang version 15.0.0 at -O3 optimization level. All the experiments are run on a MacBook Air 15.3 with an Apple M3 chip (8 cores, 2.22 GHz), 16GB RAM, and macOS 14.4.1 (23E224).

\paragraph{Experiment Results.}
Our main experiment evaluates both the error of the estimated Chamfer distance and the running time of the algorithms, over the sliding windows. 
We depict the relative error curve in \Cref{fig:exp:err} and \Cref{fig:exp:err_outlier}.

\begin{figure}[H]
\begin{subfigure}{\textwidth}
        \centering
    \includegraphics[width=\linewidth]{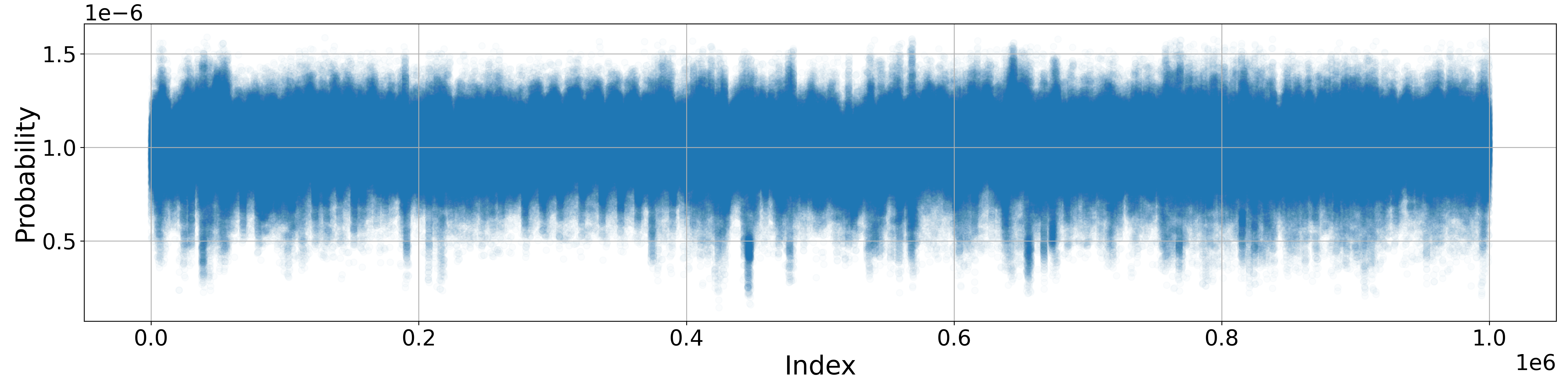}
    \end{subfigure}
    \caption{The value $\dist(a, B) / \sum_{a' \in A} \dist(a', B)$ over all points $a \in A$ for SIFT dataset,
    which are the ``ideal'' probabilities for importance sampling.}
    \label{fig:exp:probability_distribution}
\end{figure}

Here, the relative error for an estimate $\widehat E$ over the accurate Chamfer distance $E$
is defined as $\frac{|E - \widehat E|}{E}$.
Overall, our algorithm achieves less than $10\%$ error using only hundreds of samples. 
Compared with Uniform baseline, our algorithm achieves comparable error and variance for datasets without outliers, 
and shows clear advantage when the dataset has the outlier.
This showcases the robustness of our algorithm.
We observe that our algorithm performs slightly worse than Uniform in the SIFT dataset (without outliers),
but this is because the distance $\{\dist(a, B) : a \in A\}$ is very uniform (see \Cref{fig:exp:probability_distribution}),
hence uniform sampling is already the ``optimal'' sampling strategy.

\begin{figure}[H]
\begin{subfigure}{\textwidth}
        \centering
    \includegraphics[width=\linewidth]{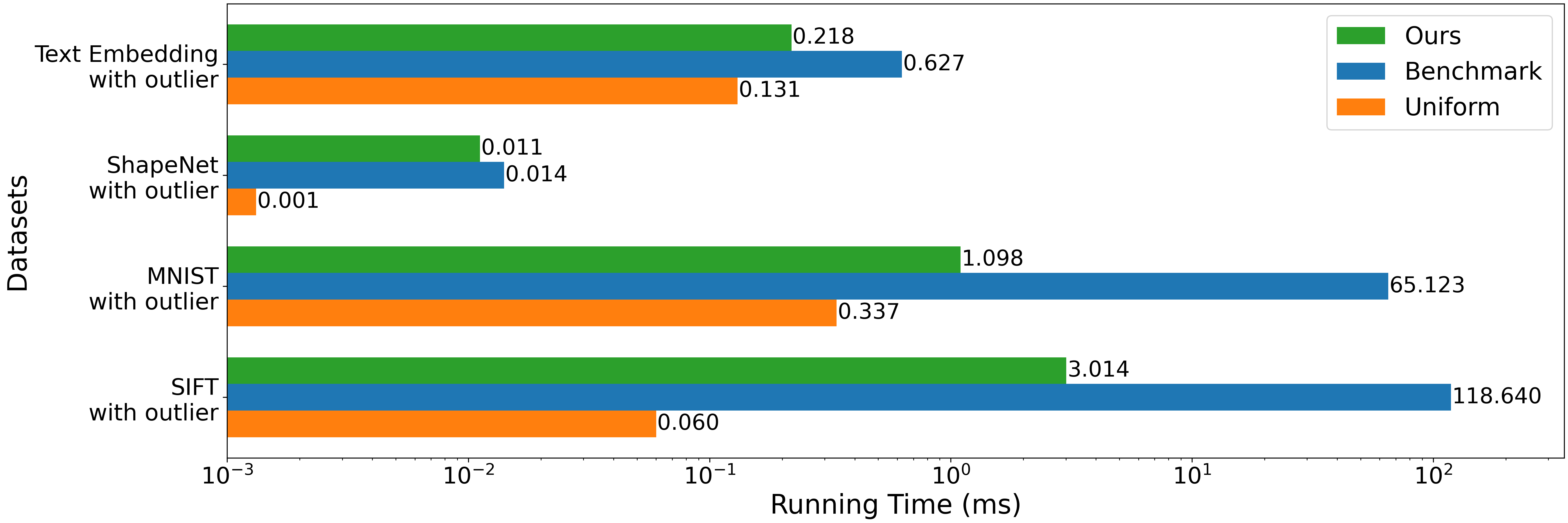}
    \end{subfigure}
    \caption{Average running time per window update for all algorithms on datasets \emph{with} outliers.}
    \label{fig:exp:run_time_outlier}
\end{figure}

We report the average running time per sliding window update in \Cref{fig:exp:run_time}.
Our algorithm is magnitudes more efficient than the Benchmark on larger datasets.
It incurs a slightly higher time cost than Uniform which is expected since Uniform does not need to maintain any additional structure to generate a sample.
The results for datasets with outliers are similar to that in \Cref{fig:exp:run_time} and can be found in \Cref{fig:exp:run_time_outlier}.


%


\begin{table}[H]
\centering
\caption{Experiment parameters when both $A$ and $B$ are dynamic.}
\begin{small}
\begin{tabular}{llllll}
\toprule
dataset  & window size & sample size\\
\midrule
Text Embedding  & 1.5k & 100\\
ShapeNet  & 1.5k & 100\\
Fashion-MNIST & 3.5k & 100\\
SIFT & 50k & 500\\
\bottomrule
\label{tab:dataset_dynamicAB}
\end{tabular}
\end{small}
\vspace{-0.2in}
\end{table}

\begin{figure}[H]
    \centering
    \begin{subfigure}{0.82\textwidth}
        \centering
    \includegraphics[width=\linewidth]{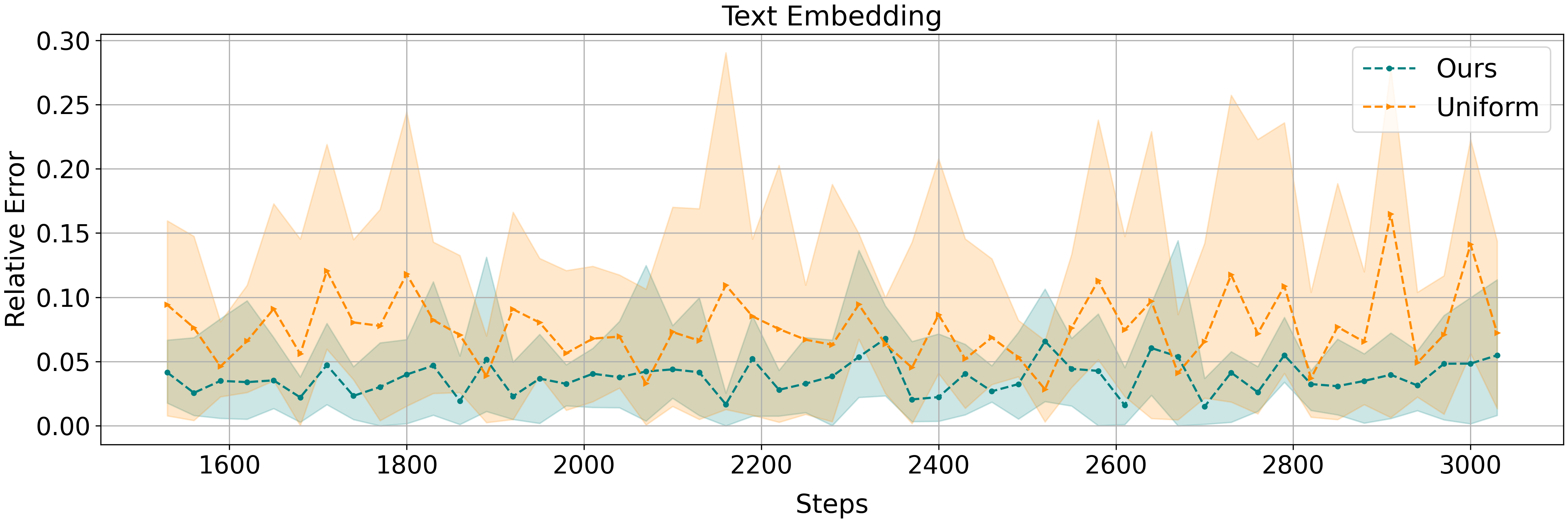}
    \end{subfigure}
    \begin{subfigure}{0.82\textwidth}
        \centering
    \includegraphics[width=\linewidth]{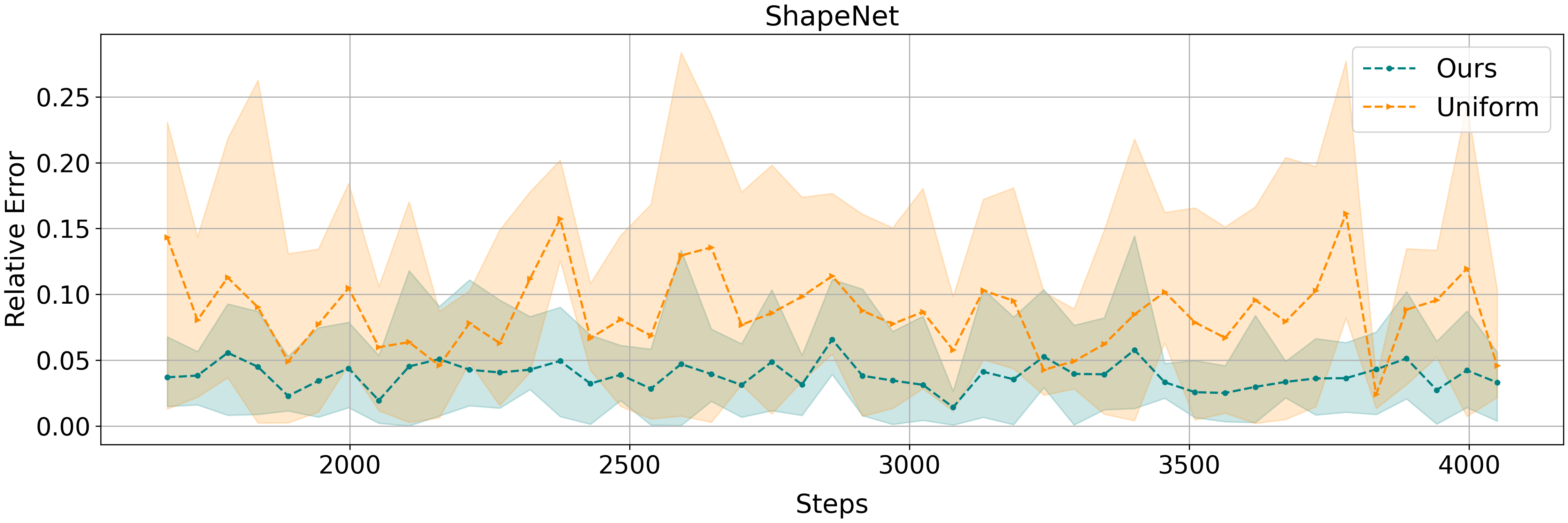}
    \end{subfigure}
    \begin{subfigure}{0.82\textwidth}
        \centering
    \includegraphics[width=\linewidth]{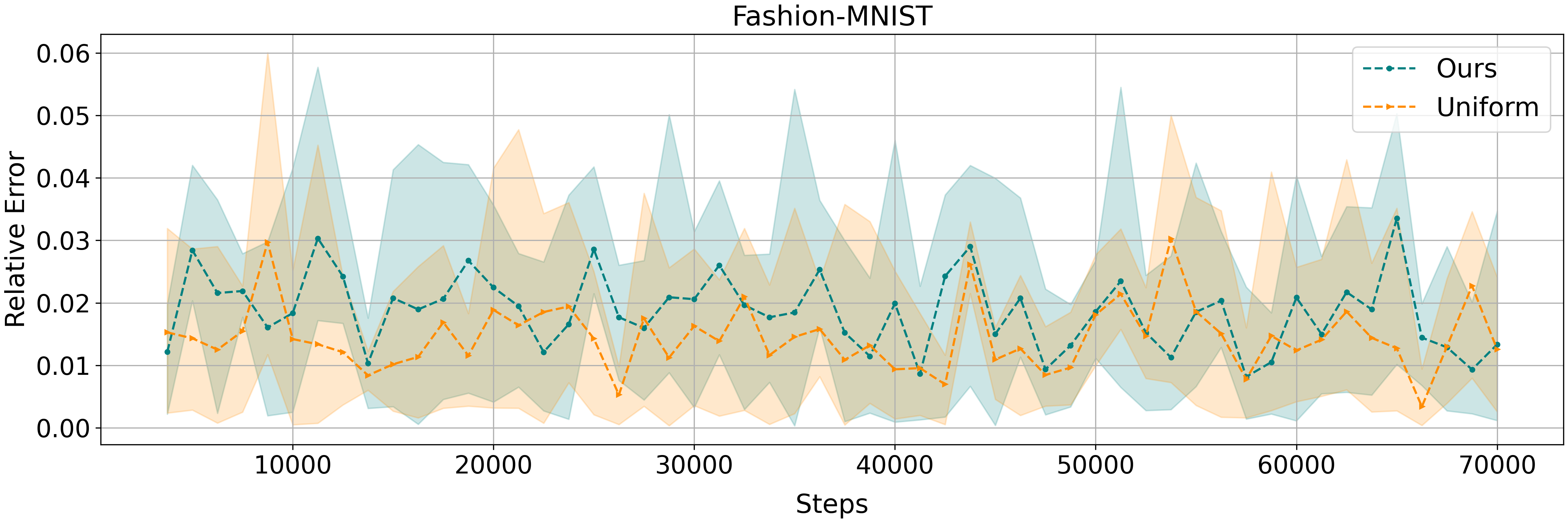}
    \end{subfigure}
    \begin{subfigure}{0.82\textwidth}
        \centering
    \includegraphics[width=\linewidth]{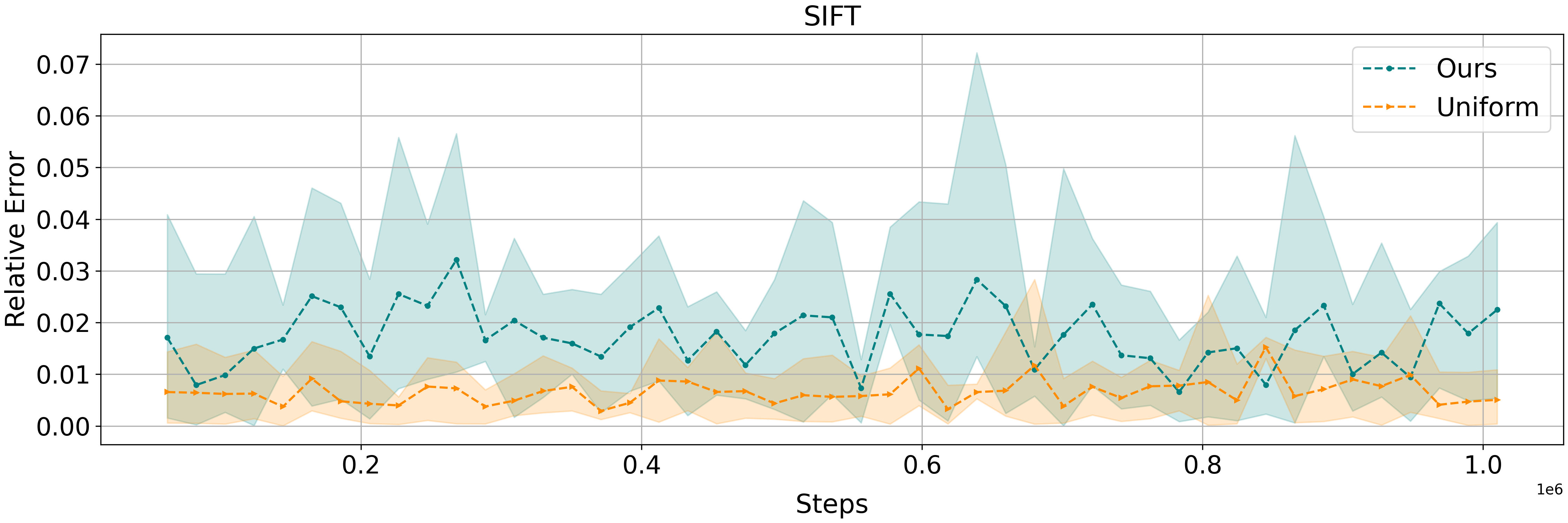}
    \end{subfigure}
    \caption{Relative error curves when both $A$ and $B$ are dynamic, with the same setup as in \Cref{fig:exp:err}.}
    \label{fig:exp:err_dynamicAB}
\end{figure}

\begin{figure}[h]
\begin{subfigure}{\textwidth}
        \centering
    \includegraphics[width=\linewidth]{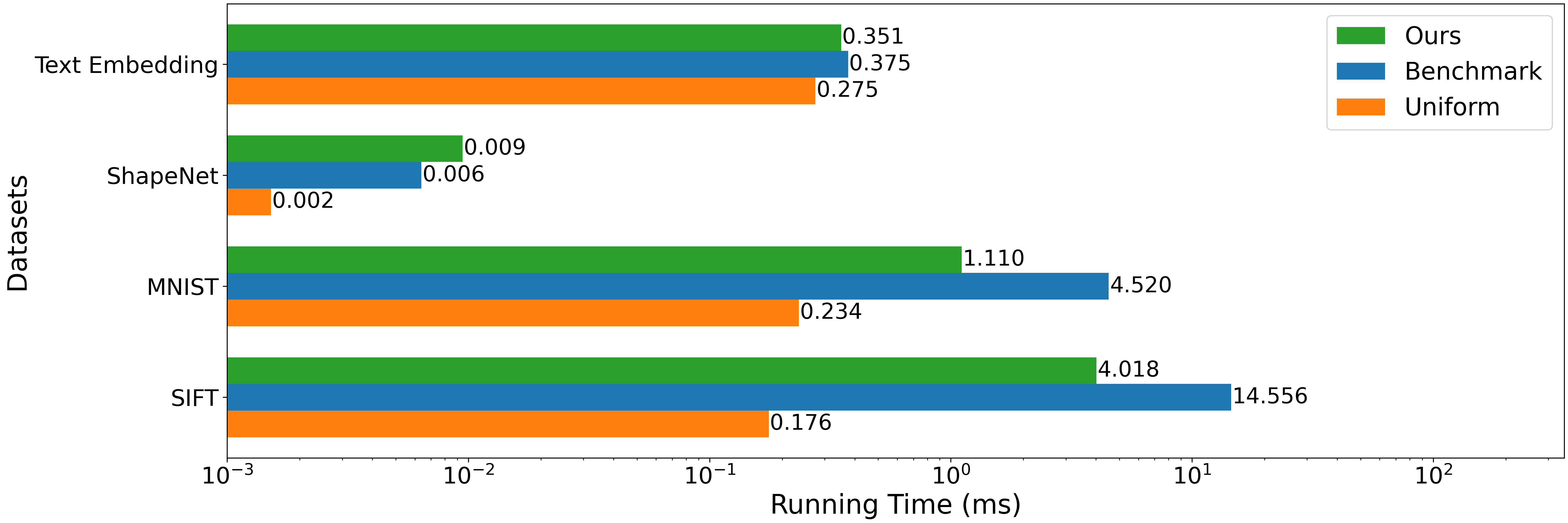}
    \end{subfigure}
    \caption{Average running time per window update for all algorithms when both $A$ and $B$ are dynamic.}
    \label{fig:exp:run_time_dynamicAB}
\end{figure}

To evaluate the case when both $A$ and $B$ are dynamic, we simulate the dynamic updates via a sliding window, similar to the default setting where only $B$ is dynamic.
However, the main difference is that the window not only consists of insertions of points in $B$, but also those in $A$.
Specifically, we insert alternatively between $A$ and $B$ in proportion to their dataset sizes (e.g., for MNIST dataset, $|A| : |B| = 6 : 1$, we perform six insertions from $A$ followed by one from $B$).
Experiment parameters for this case are summarized in \Cref{tab:dataset_dynamicAB}.
We depict the relative error curve in \Cref{fig:exp:err_dynamicAB} and report the average running time per sliding window update in \Cref{fig:exp:run_time_dynamicAB}.

\clearpage

\appendix

\section{Linear Recourse for Maintaining Assignments} \label{app:recourse}

\begin{lemma} \label{lem:highRecourse}
Any dynamic algorithm that maintains an $\alpha$-approximate assignment between $A$ and $B$ must have at least $\Omega(n)$ recourse, and thus also $\Omega(n)$ update time.   
\end{lemma}
\begin{proof}
Consider an instance where points of $A$ are in close proximity to each other, and $B$ contains two points, one at distance $1$ from $A$ and the other at distance $\Omega{(\alpha)}$. Any $\alpha$-approximate assignment of this instance must assign the majority of points in $A$ to the closer point in $B$. If the closer point of $B$ is removed from the input, then any assignment must assign all points of $A$ to the single remaining point in $B$, leading to the claimed recourse. In fact, for any constant $\delta > 0$, ~\cite{bakshi2023near} show that even in the \emph{static} setting, reporting a $(1+\epsilon)$-approximate mapping still requires $\Omega(n^{2-\delta})$ time under the hitting set conjecture~\cite{williams2018some}.
\end{proof}

\section{Our Algorithms for the $\ell_2$ Norm}

\cite{matouvsek2013lecture} has shown the following useful lemma (which we state in the dynamic setting):

\begin{lemma}
    There is a dynamic algorithm which for $n$ points $A \subset \mathbb{R}^d$ undergoing insertions and deletions in $O(d/\epsilon^2)$ worst-case update time maintains $A \subset \mathbb{R}^{O(d/\epsilon^2)}$ such that $\|A_i' - A_j'\|_1 \leq \|A_i - A_j\|_2 \leq \|A_i' - A_j'\|_1 \cdot (1+\epsilon)$ for all $i,j \in [n]$ with high ($1-1/\poly(n)$) probability.
\end{lemma}

That is, we may embed the input points into the $O(d/\epsilon^2)$ space such that $\ell_1$ distances between the embedded points roughly correspond to $\ell_2$ distances between the input points. Substituting the embedded point set into Theorem~\ref{thm:main}, and using the state of the art dynamic nearest neighbor datastructures of \cite{AryaMNSW98, AndoniR15}, we obtain the following algorithms for the $\ell_2$ norm. 
\begin{itemize}
    \item For $d = O(1)$ and $0 < \epsilon < 1$, there exists an algorithm which, with high probability, maintains a $(1+\epsilon)$-approximation to the Chamfer distance between $A,B \subset \mathbb{R}^d, |A|,|B| \leq n$ with respect to the $\ell_2$ norm, as $A$ and $B$ undergo point insertions and deletions, in $\tilde{O}(\epsilon^{-O(d/\epsilon^2)})$ worst-case update time.
    \item There exists a dynamic algorithm which maintains an $O(\epsilon^{-1})$-approximation for $0 < \epsilon < 1$ to the Chamfer distance between $A,B \subset \mathbb{R}^d, |A|,|B| \leq n$ with respect to the $\ell_2$ norm, as $A$ and $B$ undergo point insertions and deletions, in $\tilde{O}(d \cdot n^{\epsilon^2} \cdot \epsilon^{-6})$ worst-case update time.
\end{itemize}

\label{app:embeding}

\clearpage
 
\bibliographystyle{gamma}
\def\bibfont{\small}
\bibliography{Ref}

@inproceedings{chan2018fully,
  author       = {T.{-}H. Hubert Chan and
                  Arnaud Guerquin and
                  Mauro Sozio},
  editor       = {Pierre{-}Antoine Champin and
                  Fabien Gandon and
                  Mounia Lalmas and
                  Panagiotis G. Ipeirotis},
  title        = {Fully Dynamic \emph{k}-Center Clustering},
  booktitle    = {Proceedings of the 2018 World Wide Web Conference on World Wide Web,
                  {WWW} 2018},
  pages        = {579--587},
  publisher    = {{ACM}},
  year         = {2018},
  url          = {https://doi.org/10.1145/3178876.3186124},
  doi          = {10.1145/3178876.3186124},
  bibsource    = {dblp computer science bibliography, https://dblp.org}
}

@inproceedings{bhattacharya2023fully,
  author       = {Sayan Bhattacharya and
                  Mart{\'{\i}}n Costa and
                  Silvio Lattanzi and
                  Nikos Parotsidis},

  title        = {Fully Dynamic $k$-Clustering in ${\tilde{O}}(k)$ Update Time},
  booktitle    = {36th \NeurIPS},
  year         = {2023},
  url          = {http://papers.nips.cc/paper\_files/paper/2023/hash/3b7ba46201bf15e5c3935272afae50db-Abstract-Conference.html},
  bibsource    = {dblp computer science bibliography, https://dblp.org}
}

@article{henzinger2020fully,
  author       = {Monika Henzinger and
                  Sagar Kale},
  editor       = {Fabrizio Grandoni and
                  Grzegorz Herman and
                  Peter Sanders},
  title        = {Fully-Dynamic Coresets},
  booktitle    = {28th \ESA},
  series       = {LIPIcs},
  volume       = {173},
  pages        = {57:1--57:21},
  publisher    = {Schloss Dagstuhl - Leibniz-Zentrum f{\"{u}}r Informatik},
  year         = {2020},
  url          = {https://doi.org/10.4230/LIPIcs.ESA.2020.57},
  doi          = {10.4230/LIPICS.ESA.2020.57},
  bibsource    = {dblp computer science bibliography, https://dblp.org}
}

@inproceedings{feng2025even,
   author       = {Ying Feng and
                  Piotr Indyk},
  editor       = {Keren Censor{-}Hillel and
                  Fabrizio Grandoni and
                  Jo{\"{e}}l Ouaknine and
                  Gabriele Puppis},
  title        = {Even Faster Algorithm for the Chamfer Distance},
  booktitle    = {52nd \ICALP},
  series       = {LIPIcs},
  volume       = {334},
  pages        = {76:1--76:18},
  publisher    = {Schloss Dagstuhl - Leibniz-Zentrum f{\"{u}}r Informatik},
  year         = {2025},
  url          = {https://doi.org/10.4230/LIPIcs.ICALP.2025.76},
  doi          = {10.4230/LIPICS.ICALP.2025.76},
  bibsource    = {dblp computer science bibliography, https://dblp.org}
    
}

@inproceedings{cohen2019fully,
  author       = {Vincent Cohen{-}Addad and
                  Niklas Hjuler and
                  Nikos Parotsidis and
                  David Saulpic and
                  Chris Schwiegelshohn},
  editor       = {Hanna M. Wallach and
                  Hugo Larochelle and
                  Alina Beygelzimer and
                  Florence d'Alch{\'{e}}{-}Buc and
                  Emily B. Fox and
                  Roman Garnett},
  title        = {Fully Dynamic Consistent Facility Location},
  booktitle    = {32nd \NeurIPS},
  pages        = {3250--3260},
  year         = {2019},
  url          = {https://proceedings.neurips.cc/paper/2019/hash/fface8385abbf94b4593a0ed53a0c70f-Abstract.html},
  bibsource    = {dblp computer science bibliography, https://dblp.org}
}

@article{goranci2025fully,
  title = 	 {Fully Dynamic {E}uclidean Bi-Chromatic Matching in Sublinear Update Time},
  author =       {Goranci, Gramoz and Kiss, Peter and Patel, Neel and Seybold, Martin P. and Szilagyi, Eva and Zheng, Da Wei},
  booktitle = 	 {Proceedings of the 42nd International Conference on Machine Learning},
  pages = 	 {20162--20186},
  year = 	 {2025},
  editor = 	 {Singh, Aarti and Fazel, Maryam and Hsu, Daniel and Lacoste-Julien, Simon and Berkenkamp, Felix and Maharaj, Tegan and Wagstaff, Kiri and Zhu, Jerry},
  volume = 	 {267},
  series = 	 {Proceedings of Machine Learning Research},
  month = 	 {13--19 Jul},
  publisher =    {PMLR},
  url = 	 {https://proceedings.mlr.press/v267/goranci25a.html},
}

@inproceedings{cohen2024dynamic,
  author       = {Vincent Cohen{-}Addad and
                  Silvio Lattanzi and
                  Andreas Maggiori and
                  Nikos Parotsidis},
  title        = {Dynamic Correlation Clustering in Sublinear Update Time},
  booktitle    = {41st \ICML},
  publisher    = {OpenReview.net},
  year         = {2024},
  url          = {https://openreview.net/forum?id=3YG55Lbcnr},
  bibsource    = {dblp computer science bibliography, https://dblp.org}
}

@inproceedings{bhattacharya2024dynamic,
  author       = {Sayan Bhattacharya and
                  Gramoz Goranci and
                  Shaofeng H.{-}C. Jiang and
                  Yi Qian and
                  Yubo Zhang},
  title        = {Dynamic Facility Location in High Dimensional Euclidean Spaces},
  booktitle    = {41st \ICML},
  publisher    = {OpenReview.net},
  year         = {2024},
  url          = {https://openreview.net/forum?id=rucbIsWoEV},
  bibsource    = {dblp computer science bibliography, https://dblp.org}
}

@article{anand2025structural,
  author       = {Emile Anand and
                  Jan van den Brand and
                  Rose McCarty},
  title        = {The Structural Complexity of Matrix-Vector Multiplication},
  journal      = {CoRR},
  volume       = {abs/2502.21240},
  year         = {2025},
  url          = {https://doi.org/10.48550/arXiv.2502.21240},
  doi          = {10.48550/ARXIV.2502.21240},
  eprinttype    = {arXiv},
  eprint       = {2502.21240},
  bibsource    = {dblp computer science bibliography, https://dblp.org}
}

@inproceedings{bakshi2023near,
  author       = {Ainesh Bakshi and
                  Piotr Indyk and
                  Rajesh Jayaram and
                  Sandeep Silwal and
                  Erik Waingarten},
  editor       = {Alice Oh and
                  Tristan Naumann and
                  Amir Globerson and
                  Kate Saenko and
                  Moritz Hardt and
                  Sergey Levine},
  title        = {Near-Linear Time Algorithm for the Chamfer Distance},
  booktitle    = {36th \NeurIPS},
  year         = {2023},
  url          = {http://papers.nips.cc/paper\_files/paper/2023/hash/d2fe3a5711a6d488da9e9a78b84ee24c-Abstract-Conference.html},
  bibsource    = {dblp computer science bibliography, https://dblp.org}
}

@inproceedings{de2007efficient,
  author       = {Mark de Berg and
                  Herman J. Haverkort and
                  Shripad Thite and
                  Laura Toma},
  editor       = {Takeshi Tokuyama},
  title        = {I/O-Efficient Map Overlay and Point Location in Low-Density Subdivisions},
  booktitle    = {Algorithms and Computation, 18th International Symposium, {ISAAC}
                  2007},
  series       = {Lecture Notes in Computer Science},
  volume       = {4835},
  pages        = {500--511},
  publisher    = {Springer},
  year         = {2007},
  url          = {https://doi.org/10.1007/978-3-540-77120-3\_44},
  doi          = {10.1007/978-3-540-77120-3\_44},
  bibsource    = {dblp computer science bibliography, https://dblp.org}
}

@inproceedings{KusnerSKW15,
  author       = {Matt J. Kusner and
                  Yu Sun and
                  Nicholas I. Kolkin and
                  Kilian Q. Weinberger},
  editor       = {Francis R. Bach and
                  David M. Blei},
  title        = {From Word Embeddings To Document Distances},
  booktitle    = {32nd \ICML},
  series       = {{JMLR} Workshop and Conference Proceedings},
  volume       = {37},
  pages        = {957--966},
  publisher    = {JMLR.org},
  year         = {2015},
  url          = {http://proceedings.mlr.press/v37/kusnerb15.html},
  bibsource    = {dblp computer science bibliography, https://dblp.org}
}

@inproceedings{WanCLYZY019,
  author       = {Ziyu Wan and
                  Dongdong Chen and
                  Yan Li and
                  Xingguang Yan and
                  Junge Zhang and
                  Yizhou Yu and
                  Jing Liao},
  editor       = {Hanna M. Wallach and
                  Hugo Larochelle and
                  Alina Beygelzimer and
                  Florence d'Alch{\'{e}}{-}Buc and
                  Emily B. Fox and
                  Roman Garnett},
  title        = {Transductive Zero-Shot Learning with Visual Structure Constraint},
  booktitle    = {32nd \NeurIPS},
  pages        = {9972--9982},
  year         = {2019},
  url          = {https://proceedings.neurips.cc/paper/2019/hash/5ca359ab1e9e3b9c478459944a2d9ca5-Abstract.html},
  bibsource    = {dblp computer science bibliography, https://dblp.org}
}

@inproceedings{AthitsosS03,
  author       = {Vassilis Athitsos and
                  Stan Sclaroff},
  title        = {Estimating 3D Hand Pose from a Cluttered Image},
  booktitle    = {2003 {IEEE} Computer Society Conference on Computer Vision and Pattern
                  Recognition {(CVPR} 2003)},
  pages        = {432--442},
  publisher    = {{IEEE} Computer Society},
  year         = {2003},
  url          = {https://doi.org/10.1109/CVPR.2003.1211500},
  doi          = {10.1109/CVPR.2003.1211500},
  bibsource    = {dblp computer science bibliography, https://dblp.org}
}

@inproceedings{LiangJG05,
  author       = {Wei Liang and
                  Yunde Jia and
                  Cheng Ge},
  editor       = {De{-}Shuang Huang and
                  Xiao{-}Ping (Steven) Zhang and
                  Guang{-}Bin Huang},
  title        = {Visual Hand Tracking Using Nonparametric Sequential Belief Propagation},
  booktitle    = {Advances in Intelligent Computing, International Conference on Intelligent
                  Computing, {ICIC} 2005, Proceedings,
                  Part {I}},
  series       = {Lecture Notes in Computer Science},
  volume       = {3644},
  pages        = {679--687},
  publisher    = {Springer},
  year         = {2005},
  url          = {https://doi.org/10.1007/11538059\_71},
  doi          = {10.1007/11538059\_71},
  bibsource    = {dblp computer science bibliography, https://dblp.org}
}

@inproceedings{FanSG17,
  author       = {Haoqiang Fan and
                  Hao Su and
                  Leonidas J. Guibas},
  title        = {A Point Set Generation Network for 3D Object Reconstruction from a
                  Single Image},
  booktitle    = {2017 {IEEE} Conference on Computer Vision and Pattern Recognition,
                  {CVPR} 2017},
  pages        = {2463--2471},
  publisher    = {{IEEE} Computer Society},
  year         = {2017},
  url          = {https://doi.org/10.1109/CVPR.2017.264},
  doi          = {10.1109/CVPR.2017.264},
  bibsource    = {dblp computer science bibliography, https://dblp.org}
}

@inproceedings{JiangSQJ18,
  author       = {Li Jiang and
                  Shaoshuai Shi and
                  Xiaojuan Qi and
                  Jiaya Jia},
  editor       = {Vittorio Ferrari and
                  Martial Hebert and
                  Cristian Sminchisescu and
                  Yair Weiss},
  title        = {{GAL:} Geometric Adversarial Loss for Single-View 3D-Object Reconstruction},
  booktitle    = {Computer Vision - {ECCV} 2018 - 15th European Conference, Munich,
                  Germany, September 8-14, 2018, Proceedings, Part {VIII}},
  series       = {Lecture Notes in Computer Science},
  volume       = {11212},
  pages        = {820--834},
  publisher    = {Springer},
  year         = {2018},
  url          = {https://doi.org/10.1007/978-3-030-01237-3\_49},
  doi          = {10.1007/978-3-030-01237-3\_49},
  bibsource    = {dblp computer science bibliography, https://dblp.org}
}

@inproceedings{HarenstamNielsenSSAC24,
  author       = {Linus H{\"{a}}renstam{-}Nielsen and
                  Lu Sang and
                  Abhishek Saroha and
                  Nikita Araslanov and
                  Daniel Cremers},
  editor       = {Ales Leonardis and
                  Elisa Ricci and
                  Stefan Roth and
                  Olga Russakovsky and
                  Torsten Sattler and
                  G{\"{u}}l Varol},
  title        = {DiffCD: {A} Symmetric Differentiable Chamfer Distance for Neural Implicit
                  Surface Fitting},
  booktitle    = {Computer Vision - {ECCV} 2024 - 18th European Conference, Proceedings, Part {LXXIII}},
  series       = {Lecture Notes in Computer Science},
  volume       = {15131},
  pages        = {432--447},
  publisher    = {Springer},
  year         = {2024},
  url          = {https://doi.org/10.1007/978-3-031-73464-9\_26},
  doi          = {10.1007/978-3-031-73464-9\_26},
  bibsource    = {dblp computer science bibliography, https://dblp.org}
}

@inproceedings{AtasuM19,
  author       = {Kubilay Atasu and
                  Thomas Mittelholzer},
  editor       = {Kamalika Chaudhuri and
                  Ruslan Salakhutdinov},
  title        = {Linear-Complexity Data-Parallel Earth Mover's Distance Approximations},
  booktitle    = {36th \ICML},
  series       = {Proceedings of Machine Learning Research},
  volume       = {97},
  pages        = {364--373},
  publisher    = {{PMLR}},
  year         = {2019},
  url          = {http://proceedings.mlr.press/v97/atasu19a.html},
  bibsource    = {dblp computer science bibliography, https://dblp.org}
}

@inproceedings{williams2018some,
  title={On some fine-grained questions in algorithms and complexity},
  author={Williams, Virginia Vassilevska},
  booktitle={Proceedings of the international congress of mathematicians: Rio de janeiro 2018},
  pages={3447--3487},
  year={2018},
    doi = {https://doi.org/10.1142/9789813272880_0188},
  organization={World Scientific}
}

@inproceedings{LinLZHYFLZZ24,
  author       = {Fangzhou Lin and
                  Haotian Liu and
                  Haoying Zhou and
                  Songlin Hou and
                  Kazunori D. Yamada and
                  Gregory S. Fischer and
                  Yanhua Li and
                  Haichong K. Zhang and
                  Ziming Zhang},
  title        = {Loss Distillation via Gradient Matching for Point Cloud Completion
                  with Weighted Chamfer Distance},
  booktitle    = {{IEEE/RSJ} International Conference on Intelligent Robots and Systems,
                  {IROS} 2024},
  pages        = {511--518},
  publisher    = {{IEEE}},
  year         = {2024},
  url          = {https://doi.org/10.1109/IROS58592.2024.10801828},
  doi          = {10.1109/IROS58592.2024.10801828},
  bibsource    = {dblp computer science bibliography, https://dblp.org}
}

@inproceedings{WuPZWLL21,
author       = {Tong Wu and
                  Liang Pan and
                  Junzhe Zhang and
                  Tai Wang and
                  Ziwei Liu and
                  Dahua Lin},
  title        = {Density-aware Chamfer Distance as a Comprehensive Metric for Point
                  Cloud Completion},
  journal      = {35th \NeurIPS},
  volume       = {abs/2111.12702},
  year         = {2021},
  url          = {https://arxiv.org/abs/2111.12702},
  eprinttype    = {arXiv},
  eprint       = {2111.12702},
  bibsource    = {dblp computer science bibliography, https://dblp.org}
}

@inproceedings{lin2023hyperbolic,
  author       = {Fangzhou Lin and
                  Yun Yue and
                  Songlin Hou and
                  Xuechu Yu and
                  Yajun Xu and
                  Kazunori D. Yamada and
                  Ziming Zhang},
  title        = {Hyperbolic Chamfer Distance for Point Cloud Completion},
  booktitle    = {{IEEE/CVF} International Conference on Computer Vision, {ICCV} 2023,
                  Paris, France, October 1-6, 2023},
  pages        = {14549--14560},
  publisher    = {{IEEE}},
  year         = {2023},
  url          = {https://doi.org/10.1109/ICCV51070.2023.01342},
  doi          = {10.1109/ICCV51070.2023.01342},
  bibsource    = {dblp computer science bibliography, https://dblp.org}
}

@inproceedings{RunzLTMKS0ASLN20,
  author       = {Martin R{\"{u}}nz and
                  Kejie Li and
                  Meng Tang and
                  Lingni Ma and
                  Chen Kong and
                  Tanner Schmidt and
                  Ian D. Reid and
                  Lourdes Agapito and
                  Julian Straub and
                  Steven Lovegrove and
                  Richard A. Newcombe},
  title        = {FroDO: From Detections to 3D Objects},
  booktitle    = {2020 {IEEE/CVF} Conference on Computer Vision and Pattern Recognition,
                  {CVPR} 2020, Seattle, WA, USA, June 13-19, 2020},
  pages        = {14708--14717},
  publisher    = {Computer Vision Foundation / {IEEE}},
  year         = {2020},
  url          = {https://openaccess.thecvf.com/content\_CVPR\_2020/html/Runz\_FroDO\_From\_Detections\_to\_3D\_Objects\_CVPR\_2020\_paper.html},
  doi          = {10.1109/CVPR42600.2020.01473},
  bibsource    = {dblp computer science bibliography, https://dblp.org}
}

@inproceedings{TyszkiewiczMPF22,
  author       = {Michal J. Tyszkiewicz and
                  Kevis{-}Kokitsi Maninis and
                  Stefan Popov and
                  Vittorio Ferrari},
  editor       = {Shai Avidan and
                  Gabriel J. Brostow and
                  Moustapha Ciss{\'{e}} and
                  Giovanni Maria Farinella and
                  Tal Hassner},
  title        = {RayTran: 3D Pose Estimation and Shape Reconstruction of Multiple Objects
                  from Videos with Ray-Traced Transformers},
  booktitle    = {Computer Vision - {ECCV} 2022 - 17th European Conference, Tel Aviv,
                  Israel, October 23-27, 2022, Proceedings, Part {X}},
  series       = {Lecture Notes in Computer Science},
  volume       = {13670},
  pages        = {211--228},
  publisher    = {Springer},
  year         = {2022},
  url          = {https://doi.org/10.1007/978-3-031-20080-9\_13},
  doi          = {10.1007/978-3-031-20080-9\_13},
  bibsource    = {dblp computer science bibliography, https://dblp.org}
}

@inproceedings{HampaliHTMKL23,
  author       = {Shreyas Hampali and
                  Tomas Hodan and
                  Luan Tran and
                  Lingni Ma and
                  Cem Keskin and
                  Vincent Lepetit},
  title        = {In-Hand 3D Object Scanning from an {RGB} Sequence},
  booktitle    = {{IEEE/CVF} Conference on Computer Vision and Pattern Recognition,
                  {CVPR} 2023, Vancouver, BC, Canada, June 17-24, 2023},
  pages        = {17079--17088},
  publisher    = {{IEEE}},
  year         = {2023},
  url          = {https://doi.org/10.1109/CVPR52729.2023.01638},
  doi          = {10.1109/CVPR52729.2023.01638},
  bibsource    = {dblp computer science bibliography, https://dblp.org}
}

@article{van1994automatic,
  title={Automatic three-dimensional correlation of CT-CT, CT-MRI, and CT-SPECT using chamfer matching},
  author={Van Herk, Marcel and Kooy, Hanne M},
  journal={Medical physics},
  volume={21},
  number={7},
  pages={1163--1178},
  year={1994},
  publisher={Wiley Online Library}
}

@inproceedings{herman1991shape,
  author       = {Gabor T. Herman and
                  Carolyn A. Bucholtz},
  editor       = {Alan C. F. Colchester and
                  David J. Hawkes},
  title        = {Shape-Based Interpolation Using a Chamfer Distance},
  booktitle    = {Information Processing in Medical Imaging, 12th International Conference,
                  IPMI'91},
  series       = {Lecture Notes in Computer Science},
  volume       = {511},
  pages        = {314--325},
  publisher    = {Springer},
  year         = {1991},
  url          = {https://doi.org/10.1007/BFb0033762},
  doi          = {10.1007/BFB0033762},
  bibsource    = {dblp computer science bibliography, https://dblp.org}
}

@article{mathews2014segmentation,
  title={Segmentation of head from ultrasound fetal image using chamfer matching and hough transform based approaches},
  author={Mathews, M and Deepa, J and James, Tonu and Thomas, Shari},
  journal={Int. J. Eng. Res. Technol},
  volume={3},
  number={5},
  pages={1065--1068},
  year={2014}
}

@inproceedings{AndoniR15,
  author       = {Alexandr Andoni and
                  Ilya P. Razenshteyn},
  editor       = {Rocco A. Servedio and
                  Ronitt Rubinfeld},
  title        = {Optimal Data-Dependent Hashing for Approximate Near Neighbors},
  booktitle    = {47th \STOC},
  pages        = {793--801},
  publisher    = {{ACM}},
  year         = {2015},
  url          = {https://doi.org/10.1145/2746539.2746553},
  doi          = {10.1145/2746539.2746553},
  bibsource    = {dblp computer science bibliography, https://dblp.org}
}

@article{AryaMNSW98,
  author       = {Sunil Arya and
                  David M. Mount and
                  Nathan S. Netanyahu and
                  Ruth Silverman and
                  Angela Y. Wu},
  title        = {An Optimal Algorithm for Approximate Nearest Neighbor Searching Fixed
                  Dimensions},
  journal      = {J. {ACM}},
  volume       = {45},
  number       = {6},
  pages        = {891--923},
  year         = {1998},
  url          = {https://doi.org/10.1145/293347.293348},
  doi          = {10.1145/293347.293348},
  bibsource    = {dblp computer science bibliography, https://dblp.org}
}

@techreport{matouvsek2013lecture,
  title={Lecture notes on metric embeddings},
  author={Matou{\v{s}}ek, Jir{\i}},
  year={2013},
  institution={Technical report, ETH Z{\"u}rich}
}

@misc{mnist_dataset,
  author       = {Han Xiao and
                  Kashif Rasul and
                  Roland Vollgraf},
  title        = {Fashion-MNIST: a Novel Image Dataset for Benchmarking Machine Learning
                  Algorithms},
  journal      = {CoRR},
  volume       = {abs/1708.07747},
  year         = {2017},
  url          = {http://arxiv.org/abs/1708.07747},
  eprinttype    = {arXiv},
  eprint       = {1708.07747},
  bibsource    = {dblp computer science bibliography, https://dblp.org}
}

@article{JegouDS11,
  author       = {Herv{\'{e}} J{\'{e}}gou and
                  Matthijs Douze and
                  Cordelia Schmid},
  title        = {Product Quantization for Nearest Neighbor Search},
  journal      = {{IEEE} Trans. Pattern Anal. Mach. Intell.},
  volume       = {33},
  number       = {1},
  pages        = {117--128},
  year         = {2011},
  url          = {https://doi.org/10.1109/TPAMI.2010.57},
  doi          = {10.1109/TPAMI.2010.57},
  bibsource    = {dblp computer science bibliography, https://dblp.org}
}

@inproceedings{BiabaniHM023,
  author       = {Leyla Biabani and
                  Annika Hennes and
                  Morteza Monemizadeh and
                  Melanie Schmidt},
  editor       = {Alice Oh and
                  Tristan Naumann and
                  Amir Globerson and
                  Kate Saenko and
                  Moritz Hardt and
                  Sergey Levine},
  title        = {Faster Query Times for Fully Dynamic k-Center Clustering with Outliers},
  booktitle    = {36th \NeurIPS},
  year         = {2023},
  url          = {http://papers.nips.cc/paper\_files/paper/2023/hash/1d8e261c241aa72f9b4a02af7f52587e-Abstract-Conference.html},
  bibsource    = {dblp computer science bibliography, https://dblp.org}
}

@article{ChanLSW24,
  author       = {T.{-}H. Hubert Chan and
                  Silvio Lattanzi and
                  Mauro Sozio and
                  Bo Wang},
  title        = {Fully Dynamic $k$-Center Clustering with Outliers},
  journal      = {Algorithmica},
  volume       = {86},
  number       = {1},
  pages        = {171--193},
  year         = {2024},
  url          = {https://doi.org/10.1007/s00453-023-01159-3},
  doi          = {10.1007/S00453-023-01159-3},
  bibsource    = {dblp computer science bibliography, https://dblp.org}
}

@inproceedings{GoranciHL18,
  author       = {Gramoz Goranci and
                  Monika Henzinger and
                  Dariusz Leniowski},
  editor       = {Yossi Azar and
                  Hannah Bast and
                  Grzegorz Herman},
  title        = {A Tree Structure For Dynamic Facility Location},
  booktitle    = {26th \ESA},
  series       = {LIPIcs},
  volume       = {112},
  pages        = {39:1--39:13},
  publisher    = {Schloss Dagstuhl - Leibniz-Zentrum f{\"{u}}r Informatik},
  year         = {2018},
  url          = {https://doi.org/10.4230/LIPIcs.ESA.2018.39},
  doi          = {10.4230/LIPICS.ESA.2018.39},
  bibsource    = {dblp computer science bibliography, https://dblp.org}

}

@inproceedings{Lacki2024,
  author       = {Jakub Lacki and
                  Bernhard Haeupler and
                  Christoph Grunau and
                  Rajesh Jayaram and
                  V{\'{a}}clav Rozhon},
  editor       = {David P. Woodruff},
  title        = {Fully Dynamic Consistent \emph{k}-Center Clustering},
  booktitle    = {\SODA},
  pages        = {3463--3484},
  publisher    = {{SIAM}},
  year         = {2024},
  url          = {https://doi.org/10.1137/1.9781611977912.124},
  doi          = {10.1137/1.9781611977912.124},
  bibsource    = {dblp computer science bibliography, https://dblp.org}
}

@inproceedings{Cruciani2024,
  author       = {Emilio Cruciani and
                  Sebastian Forster and
                  Gramoz Goranci and
                  Yasamin Nazari and
                  Antonis Skarlatos},
  editor       = {David P. Woodruff},
  title        = {Dynamic algorithms for \emph{k}-center on graphs},
  booktitle    = {\SODA},
  pages        = {3441--3462},
  publisher    = {{SIAM}},
  year         = {2024},
  url          = {https://doi.org/10.1137/1.9781611977912.123},
  doi          = {10.1137/1.9781611977912.123},
  bibsource    = {dblp computer science bibliography, https://dblp.org}
}

@inproceedings{BateniEFHJMW23,
  author       = {MohammadHossein Bateni and
                  Hossein Esfandiari and
                  Hendrik Fichtenberger and
                  Monika Henzinger and
                  Rajesh Jayaram and
                  Vahab Mirrokni and
                  Andreas Wiese},
  editor       = {Nikhil Bansal and
                  Viswanath Nagarajan},
  title        = {Optimal Fully Dynamic \emph{k}-Center Clustering for Adaptive and
                  Oblivious Adversaries},
  booktitle    = {\SODA},
  pages        = {2677--2727},
  publisher    = {{SIAM}},
  year         = {2023},
  url          = {https://doi.org/10.1137/1.9781611977554.ch101},
  doi          = {10.1137/1.9781611977554.CH101},
  bibsource    = {dblp computer science bibliography, https://dblp.org}
}

@inproceedings{GoranciHLSS21,
  author       = {Gramoz Goranci and
                  Monika Henzinger and
                  Dariusz Leniowski and
                  Christian Schulz and
                  Alexander Svozil},
  editor       = {Martin Farach{-}Colton and
                  Sabine Storandt},
  title        = {Fully Dynamic \emph{k}-Center Clustering in Low Dimensional Metrics},
  booktitle    = {Proceedings of the Symposium on Algorithm Engineering and Experiments,
                  {ALENEX} 2021},
  pages        = {143--153},
  publisher    = {{SIAM}},
  year         = {2021},
  url          = {https://doi.org/10.1137/1.9781611976472.11},
  doi          = {10.1137/1.9781611976472.11},
  bibsource    = {dblp computer science bibliography, https://dblp.org}
}

@inproceedings{DBLP:conf/nips/Cohen-AddadHPSS19,
  author       = {Vincent Cohen{-}Addad and
                  Niklas Hjuler and
                  Nikos Parotsidis and
                  David Saulpic and
                  Chris Schwiegelshohn},
  editor       = {Hanna M. Wallach and
                  Hugo Larochelle and
                  Alina Beygelzimer and
                  Florence d'Alch{\'{e}}{-}Buc and
                  Emily B. Fox and
                  Roman Garnett},
  title        = {Fully Dynamic Consistent Facility Location},
  booktitle    = {32nd \NeurIPS},
  pages        = {3250--3260},
  year         = {2019},
  url          = {https://proceedings.neurips.cc/paper/2019/hash/fface8385abbf94b4593a0ed53a0c70f-Abstract.html},
  bibsource    = {dblp computer science bibliography, https://dblp.org}
}

@article{DBLP:journals/corr/ChangFGHHLSSSSX15,
  author       = {Angel X. Chang and
                  Thomas A. Funkhouser and
                  Leonidas J. Guibas and
                  Pat Hanrahan and
                  Qi{-}Xing Huang and
                  Zimo Li and
                  Silvio Savarese and
                  Manolis Savva and
                  Shuran Song and
                  Hao Su and
                  Jianxiong Xiao and
                  Li Yi and
                  Fisher Yu},
  title        = {ShapeNet: An Information-Rich 3D Model Repository},
  journal      = {CoRR},
  volume       = {abs/1512.03012},
  year         = {2015},
  url          = {http://arxiv.org/abs/1512.03012},
  eprinttype    = {arXiv},
  eprint       = {1512.03012},
  bibsource    = {dblp computer science bibliography, https://dblp.org}
}

\end{document}